\documentclass[a4paper,11pt]{article}
\pdfoutput=1

\usepackage{amsfonts}
\usepackage{amsmath}
\usepackage{amssymb}
\usepackage{mathrsfs}
\usepackage{amsthm}
\usepackage[english]{babel}
\usepackage[latin1]{inputenc}
\usepackage[T1]{fontenc}
\usepackage{epsfig}
\usepackage{textcomp}
\usepackage{graphicx}
\usepackage{epstopdf}
\usepackage{subfigure}
\usepackage{dcolumn}
\usepackage{bm}
\usepackage{booktabs}

\numberwithin{equation}{section}
\def\a{\alpha}

\def\g{\gamma}

\def\d{\delta}
\def\D{\Delta}

\def\z{\zeta}

\def\th{\theta}

\def\la{\lambda}

\def\m{\mu}
\def\n{\nu}

\def\r{\rho}

\def\vf{\varphi}

\def\o{\omega}
\def\O{\Omega}

\providecommand{\abs}[1]{\lvert#1\rvert}

\newcommand{\de}{\partial}

\newcommand{\p}{\prime}
\newcommand{\beq}{\begin{equation}}
\newcommand{\eeq}{\end{equation}}

\newcommand{\noi}{\noindent}

\newcommand{\mcal}[1]{\mathcal{#1}}
\newcommand{\mscr}[1]{\mathscr{#1}}

\newcommand{\mbbR}{\mathbb{R}}

\begin{document}

\date{\mbox{}}

\title{
{\bf \huge Neutron star masses in \boldmath $R^{2}$-gravity}
 \\[8mm]
}

\author{
Fulvio Sbis\`a$^{1}$\thanks{fulviosbisa@gmail.com (corresponding author)} \hspace{1pt}, Pedro O.\ Baqui$^{1}$\thanks{pedrobaqui@gmail.com} \hspace{1pt}, Tays Miranda$^{1,2}$\thanks{tays.miranda@cosmo-ufes.org} \hspace{1pt}, Sergio E.\ Jor\'as$^{3}$\thanks{joras@if.ufrj.br} \hspace{3pt} \\[2mm]
and Oliver F.\ Piattella$^{1,4}$\thanks{oliver.piattella@cosmo-ufes.org}
\\[8mm]
\normalsize\it
$^1$ Departamento de F\'isica, Universidade Federal do Esp\'irito Santo,\\
\normalsize\it
Avenida Fernando Ferrari, 514, CEP 29075-910, Vit\'oria, ES, Brazil \vspace{.5cm} \\
\normalsize\it
$^2$ Institute of Cosmology \& Gravitation, University of Portsmouth,\\
\normalsize\it
Dennis Sciama Building, Portsmouth, PO1 3FX, United Kingdom \vspace{.5cm} \\
\normalsize\it
$^3\,$ Instituto de F\'isica, Universidade Federal do Rio de Janeiro,\\
\normalsize\it
C.P.\ 68528, CEP 21941-972, Rio de Janeiro, RJ, Brazil \vspace{.5cm} \\
\normalsize\it
$^4\,$ Institut f\"ur Theoretische Physik, Ruprecht-Karls-Universit\"at Heidelberg,\\
\normalsize\it
Philosophenweg 16, 69120 Heildelberg, Germany \\
}

\maketitle

\setcounter{page}{1}
\thispagestyle{empty}

\begin{abstract}
We address the issue of the existence of inequivalent def\mbox{}initions of gravitational mass in $R^{2}$-gravity. We present several def\mbox{}initions of gravitational mass, and discuss the formal relations between them. We then consider the concrete case of a static and spherically symmetric neutron star, and solve numerically the equations of motion for several values of the free parameter of the model. We compare the features of the mass-radius relations obtained for each def\mbox{}inition of gravitational mass, and we comment on their dependence on the free parameter. We then argue that $R^{2}$-gravity is a valuable proxy to discuss the existence of inequivalent def\mbox{}initions of gravitational mass in a generic modif\mbox{}ied gravity theory, and present some comments on the general case.
\end{abstract}

\smallskip

\noi Keywords: modif\mbox{}ied gravity, f(R) gravity, neutron stars.

\section{Introduction}
\label{sec: intro}

Infrared modif\mbox{}ications of gravity have become a very popular way of addressing the problem of the late-time acceleration of the Universe. The possibility of explaining the cosmic acceleration without introducing exotic and experimentally unobserved forms of energy, has generated a lot of interest in theories like $f(R)$, braneworlds, massive gravity and generalisations thereof. Currently, they are undergoing an intense scrutiny as a result of the recent birth of the f\mbox{}ield of multimessenger astronomy.

A notoriously delicate aspect of modif\mbox{}ied theories of gravity (MTG) is that, since modifying General Relativity (GR) introduces additional degrees of freedom (DsOF) in the theory, often the late time acceleration is achieved at the expense of the correct behaviour of gravity at small scales. Indeed, an ef\mbox{}f\mbox{}icient screening of the additional DsOF is mandatory to ensure that a MTG passes the solar system tests. The physics of compact objects provides another important test bench, this time regarding the strong gravity regime. Changing the behaviour of gravity in fact generally impacts both on the macroscopic properties of massive bodies and on their evolution history.

A compelling case is that of neutron stars (NS). In the static and spherically symmetric case GR makes rather stark predictions, like the existence of a minimum and a maximum mass for NS. This implies that, provided one is able to account for the ef\mbox{}fects of rotation and to model reliably their internal structure, observations of NS can be used to test (and potentially falsify) GR. As a matter of fact, observations like PSR J1614--2230 \cite{Demorest 2010} and PSR J0348--0432 \cite{Antoniadis 2013} already represent a challenge for GR. However, a satisfactory understanding of the behaviour of nuclear matter at the extreme conditions realised in the interior of a NS is still lacking, so it is hard to say whether these observations point to a breakdown of GR or to a poor understanding of the equation of state (EoS). An important tool to break this degeneracy is the mass-radius (M--R) relation. The interest in this f\mbox{}ield is presently very high due to the recent detection of gravitational wave signals by the LIGO and VIRGO collaborations, exemplif\mbox{}ied by the signal GW 170817 emitted by a merging neutron stars binary system \cite{TheLIGOScientific:2017qsa, Abbott:2017xzu, Monitor:2017mdv, TheLIGOScientific:2016src, LIGOScientific:2019fpa, Abbott:2018wiz, Abbott:2018exr}.

However, a subtle and perhaps underestimated point is that, when we speak of the mass of a star in MTG, we are speaking of a not well-def\mbox{}ined concept. In GR we are used to consider dif\mbox{}ferent def\mbox{}initions of mass, depending on the specif\mbox{}ic problem under consideration, which are nonetheless equivalent. A crucial remark is that these def\mbox{}initions are in general \emph{not} equivalent in MTG, when additional degrees of freedom enter into play, so possible confusion arises when we speak generically of ``gravitational mass'' in MTG without specifying to what def\mbox{}inition exactly we are referring to. Clarifying this point is evidently very important, especially to be able to compare dif\mbox{}ferent types of observations, theoretical predictions with observations, and dif\mbox{}ferent theoretical studies (see on this respect the case of \cite{Yazadjiev:2014cza, Yazadjiev:2015xsj} and \cite{Capozziello:2015yza}, which use dif\mbox{}ferent def\mbox{}initions of mass and f\mbox{}ind dif\mbox{}ferent M--R relations).

Our aim here is to give a systematic discussion of this fact. Although the inequivalence is, as we propose, inherent to MTG, considering a general MTG from the outset would imply the risk of not being able to identify clearly the physical reason for the inequivalence. To avoid generality to hinder clarity of analysis, we prefer to consider a concrete model of MTG, to be used as a proxy which is meant to represent general features of MTG. For this reason we focus on the quadratic $f(R)$ model $f(R) = R + \a \, R^2$ (to which we refer to as $R^{2}$-gravity), also known as the Starobinsky model \cite{Starobinsky:1980te, starobinskynonsingular} (which is one of the most successful models in inf\mbox{}lationary cosmology \cite{Martin:2013nzq}). Coherently with the spirit of our analysis of using this model as a proxy, we leave $\alpha$ completely free, although it is known that its value is observationally severely constrained both from cosmological observations \cite{Amendola:2006we, Akrami:2018odb} and from solar system tests \cite{Bertotti:2003rm, Will:2014kxa}. For the sake of concreteness, and to be able to grasp the quantitative relevance of the inequivalence, we apply our discussion to the M--R relation of neutron stars in $R^{2}$-gravity, investigating how the features of the curves change when we change the def\mbox{}inition of gravitational mass.

The paper is structured as follows: in Section~\ref{sec: equations of motion} we review the equations of motion in $R^2$-gravity, specialising to a static and spherically symmetric system. In Section~\ref{sec: definition of mass} we introduce several def\mbox{}initions of gravitational mass, discussing their inequivalence. In Section~\ref{sec: numerical mass radius} we consider a neutron star, and study its M--R relation using the def\mbox{}initions of gravitational mass previously introduced. We discuss our results and present our conclusions in Section~\ref{sec: Discussion and conclusions}.

We adopt the ``mostly plus'' signature $(-,+,+,+)$ and, unless stated otherwise, we use units of measure where $c = 1\,$.

\section{The equations of motion and the behaviour of weak gravity}
\label{sec: equations of motion}

The general $f(R)$ gravity action \cite{Sotiriou:2008rp, DeFelice:2010aj, Nojiri:2010wj, Nojiri:2017ncd} is given by
\begin{equation} \label{general action}
	S = \frac{1}{16 \pi G_{_{\! N}}} \, \int d^{4}x \, \sqrt{-g} \,\, f(R) + S_{_{\! M}} \quad ,
\end{equation}
where $g$ is the determinant of the metric $g_{\m\n}$, $R$ is the Ricci scalar and $S_{_{\! M}}$ is the action of matter f\mbox{}ields. The case of $R^{2}$-gravity corresponds to the choice
\begin{equation} \label{Starobinsky model}
	f(R) = R + \a \, R^{2} \quad ,
\end{equation}
with $\a > 0\,$. In the metric formulation, which we adopt throughout the paper, the connection is taken to be the Levi-Civita one and the metric obeys the equation of motion
\begin{equation}\label{metric eq}
	\big( 1 + 2 \a R \big) \, R_{\m\n} - \frac{1}{2} \, g_{\m\n} \, \big( R + \alpha R^2 \big) - 2 \a \, \big( \nabla_{\!\m} \nabla_{\!\n} - g_{\m\n} \, \Box \big) \, R = 8 \pi G_{_{\! N}} \, T_{\m\n} \quad ,
\end{equation}
where the Ricci scalar is a functional of the metric $R = R \, [g_{\m\n}]\,$, $\Box = g^{\m\n} \, \nabla_{\!\m} \, \nabla_{\!\n}$ is the (curved space) d'Alembert operator and the stress-energy tensor is def\mbox{}ined as
\begin{equation} \label{stress-energy tensor}
	T_{\m\n} = - \frac{2} {\sqrt{-g}} \frac{\d S_{_{\! M}}}{\d g^{\m\n}} \quad .
\end{equation}

As is well-known, the theory can be recast in a scalar-tensor form. Introducing the scalar degree of freedom
\begin{equation} \label{scalar dof}
	\z = \a R \quad ,
\end{equation}
the fourth-order equation (\ref{metric eq}) can be shown to be equivalent to the second-order system for the metric $g_{\m\n}$ and the scalar f\mbox{}ield $\z$ \cite{Teyssandier:1983zz}:
\begin{align}
	\big( 1 + 2 \z \big) \, G_{\m}{}^{\n} &= 2 \, \Big( g^{\n\la} \nabla_{\m} \, \de_{\la} - \d_{\m}{}^{\n} \, \Box \Big) \z - 3 m^2 \, \z^2 \, \delta_{\m}{}^{\n} + 8 \pi G_{_{\! N}} \, T_{\m}{}^{\n} \quad , \label{metric eq gen} \\[3mm]
	\Box \, \z - m^2 \z &= \frac{4 \pi G_{_{\! N}}}{3} \, T \quad , \label{zeta eq gen}
\end{align}
where $T$ is the trace of the stress-energy tensor and we def\mbox{}ined the mass associated to $\z$ as
\begin{equation} \label{mass of zeta}
	m = \frac{1}{\sqrt{6 \a}} \quad .
\end{equation}
Although linked by the relation (\ref{scalar dof}), the metric and the scalar f\mbox{}ield are independent degrees of freedom as far as the initial value problem of the system (\ref{metric eq gen})--(\ref{zeta eq gen}) is concerned. It is possible to perform a f\mbox{}ield redef\mbox{}inition to avoid having $(1 + 2 \z)$ multiply $G_{\m\n}$, at the expense of introducing a non-minimal coupling with matter (Einstein frame). See e.g.\ \cite{Kase:2019dqc} for an analysis in that direction. In this work, we don't follow that path and work only in the Jordan frame.

\subsection{Non-rotating stars}
\label{subsec: static stars}

To study static and spherically symmetric stars in $R^{2}$-gravity, we consider the line element
\begin{equation} \label{metric}
	ds^{2} = - B(r) \, dt^{2} + A(r) \, dr^{2} + r^{2} \, \big( d \th^{2} + \sin^{2}\! \th \, d\vf^{2} \big) \quad ,
\end{equation}
and model the star as a perfect f\mbox{}luid, so the stress-energy tensor reads
\begin{equation} \label{stress energy st sph}
	T_{\m}^{\,\,\, \n} = diag \, \big( - \r \, , p \, , p \, , p \, \big) \quad ,
\end{equation}
where $\rho(r)$ and $p(r)$ are the energy density and the pressure, respectively. We postpone to Section \ref{sec: numerical mass radius} the detailed discussion of the equation of state we employ to model the interior of a neutron star.

The staticity condition implies that the covariant conservation of $T_{\m}{}^{\n}$ gives the equation of hydrostatic equilibrium
\begin{equation} \label{hydrostatic eq}
	p^{\p} = -\frac{B^{\p}}{2 B} \, \big( \rho + p \, \big)  \quad ,
\end{equation}
while the equation (\ref{zeta eq gen}) for the scalar DOF takes the form
\begin{equation} \label{zeta eq}
	\z^{\p\p} + \left( \frac{2}{r} + \frac{B^{\p}}{2 B} - \frac{A^{\p}}{2 A} \right) \, \z^{\p} = A \, \bigg[ \, m^{2} \z + \frac{4 \pi G_{_{\! N}}}{3} \, \big( 3p - \r \big) \, \bigg] \quad ,
\end{equation}
where we indicated $\phantom{i}^{\p} \equiv d/dr \,$. By taking suitable linear combinations of the components of (\ref{metric eq gen}), we obtain the equations for $B$ and $A$
\begin{subequations}\label{systemTOV}
\begin{align}
	&\Big( 1 + 2 \z + r \z^{\p} \Big) \, \frac{1}{A r} \frac{B^{\p}}{B} = \frac{1 + 2 \z}{r^2} \, \bigg(1 - \frac{1}{A} \bigg) - 3 m^2 \z^{2} - \frac{4}{A r} \, \z^{\p} + 8 \pi G_{_{\! N}} \, p \quad , \label{Aeqfull} \\[5mm]
	&\Big( 1 + 2 \z + r \z^{\p} \Big) \, \frac{1}{A r} \frac{A^{\p}}{A} = \frac{1 + 2 \z}{r^2} \, \bigg( \frac{1}{A} - 1 \bigg) + 3 m^2 \z^{2} + \frac{2}{A} \, \bigg( \z^{\p\p} + \frac{2}{r} \, \z^{\p} \bigg) + 8 \pi G_{_{\! N}} \, \rho \quad . \label{Beqfull}
\end{align}
\end{subequations}

\subsection{The behaviour of weak gravity}
\label{subsec: behaviour of weak gravity}

In light of the discussion to follow, it is worthwhile to recall the features of static and weak gravitational f\mbox{}ields in $R^{2}$-gravity. Although non-linear ef\mbox{}fects do become important for neutron stars, the results of this subsection will be useful for the discussion of Section \ref{sec: definition of mass} on the concept of gravitational mass. To prevent any possible confusion we underline that, when studying neutron stars in Section \ref{sec: numerical mass radius}, we consider the full non-linear equations of motion.

The behaviour of the gravitational f\mbox{}ield outside a static and spherically symmetric star in $f(R)$ gravity has been extensively studied in the literature. We follow here the analysis of \cite{Sbisa:2018rem}, where an exhaustive list of references can be found.\footnote{Note however that, to facilitate the comparison with \cite{Resco:2016upv, Astashenok:2017dpo}, we changed notation calling now $B$ and $A$ respectively what we called $A$ and $B$ in \cite{Sbisa:2018rem}.} In particular, we consider the spacetime to be asymptotically f\mbox{}lat, which is a consistent boundary condition when the function $f(R)$ is analytic in $R = 0\,$. The behaviour of the metric and of the scalar DOF is controlled by two characteristic radii $r_{\!\z}$ and $r_{\!g}$, which somehow play in $R^{2}$-gravity a role analogous to that of the Schwarzschild radius in GR. Indicating with $r_{\star}$ the radius of the star, and slightly changing the notation with respect to \cite{Sbisa:2018rem}, the characteristic radii take the form
\begin{subequations} \label{characteristic radii}
\begin{align}
	r_{\!\z} &= 2 G_{_{\! N}} \Big( \tilde{M}_{\r} - 3 \tilde{P} \Big) \quad , \label{r zeta} \\[3mm]
	r_{\!g} &= 2 G_{_{\! N}} \Bigg[ \bigg( \frac{\tilde{M}_{\r}}{3} - \tilde{P} \bigg) \, e^{-mr_\star} \, \big( 1 + mr_\star \big) + \frac{2 M_{\r}}{3} + P + \Xi \, \Bigg] \quad , \label{r g}
\end{align}
\end{subequations}
where
\begin{subequations}
\begin{align}
	M_{\r} &= 4 \pi \int_{0}^{r_\star} \!\! \rho(r) \, r^2 dr \quad , & P &= 4 \pi \int_{0}^{r_\star} \!\! p(r) \, r^2 dr \quad , \label{M and P def} \\[3mm]
	\tilde{M}_{\r} &= 4 \pi \int_{0}^{r_\star} \frac{\sinh mr}{mr} \, \rho(r) \, r^2 dr \quad , & \tilde{P} &= 4 \pi \int_{0}^{r_\star} \frac{\sinh mr}{mr} \, p(r) \, r^2 dr \quad , \label{M and P tilde def} \\[3mm]
	\Xi &= \frac{m^{2}}{G_{_{\! N}}} \int_{0}^{r_{\star}} \! \z(r) \, r^2 dr \quad . \label{Xi def}
\end{align}
\end{subequations}
Note that $\tilde{M}_{\r} \to M_{\r}\,$, $\tilde{P} \to P \,$ and $\Xi \to 0$ when $m \to 0\,$.

In the weak-f\mbox{}ield approximation, the solutions outside of the star explicitly read \cite{Sbisa:2018rem}
\begin{subequations} \label{sol ext}
\begin{align}
	\z(r) &= \frac{r_{\!\z}}{6 r} \, e^{-mr} \quad , \label{zeta sol ext} \\[4mm]
	B(r) &= 1 - \frac{r_{\!\z}}{3 r} \, e^{-mr} - \frac{r_{\!g}}{r} \quad , \label{A sol ext} \\[3mm]
	\frac{1}{A(r)} &= 1 + \big( 1 + mr \big) \, \frac{r_{\!\z}}{3 r} \, e^{-mr} - \frac{r_{\!g}}{r} \quad . \label{B sol ext}
\end{align}
\end{subequations}
It is apparent that, at distances $r \gg m^{-1}$ larger than the range of the scalar DOF, the Schwarzschild solution is recovered with $r_{\! g}$ as the ef\mbox{}fective (asymptotic) Schwarzschild radius. In this region, to which we refer to as the \emph{asymptotic region}, GR is recovered. On the other hand, at distances $r \sim m^{-1}$ comparable to the range of the scalar DOF, the metric components do not even have a Newtonian behaviour, due to the presence of the exponential function. We refer to this region as the \emph{transition region}. Finally, when the range of the scalar DOF is much larger than the radius of the star ($r_\star \ll m^{-1}$), there exist another interesting region $r_\star \leq r \ll m^{-1}$ outside the star, to which we refer to as the \emph{nearby region}. In this region $\z$, $B$ and $A$ display a Newtonian behaviour
\begin{subequations} \label{sol ext nearby}
\begin{align}
	\z(r) &\simeq \frac{r_{\!\z}}{6 r} \quad , \label{zeta sol nearby} \\[4mm]
	B(r) &\simeq 1 - \frac{r_{\!\z} + 3 r_{\!g}}{3 r} \quad , \label{A sol nearby} \\[3mm]
	\frac{1}{A(r)} &\simeq 1 + \frac{r_{\!\z} - 3 r_{\!g}}{3 r} \quad , \label{B sol nearby}
\end{align}
\end{subequations}
but the gravitational potentials have dif\mbox{}ferent amplitude, since they are sourced in a dif\mbox{}ferent way by the scalar DOF. More precisely, the PPN parameter $\g$ in the nearby region reads \cite{Sbisa:2018rem}
\begin{equation} \label{gamma nearby}
	\gamma_{\star} \simeq \frac{2 M_{\r} + 3 P}{4 M_{\r} - 3 P} \quad ,
\end{equation}
where neglecting the pressure we recover the well-known result $\gamma_{\star} = 1/2$ \cite{Chiba:2003ir}. Clearly GR is not recovered here, apart from the hypothetical case where exotic physics inside the star pushes the pressure to be of the same order of the mass, since solar system observations constrain $\g$ to be unity within few parts in $10^5$ \cite{Will:2014kxa}. Nevertheless, if we limit our attention to non-relativistic bodies, Newtonian gravity is recovered.

\section{On the def\mbox{}inition of gravitational mass in \boldmath $R^{2}$-gravity}
\label{sec: definition of mass}

To put the discussion of the inequivalent def\mbox{}initions of gravitational mass in the proper context, and to spell out the subtleties which become relevant in the $R^2$ case, it is worthwhile to recall f\mbox{}irst how the concept of gravitational mass emerges in Newtonian gravity and in GR.

\subsection{The gravitational mass in Newton's and Einstein's theories}

In Newtonian gravity, the concept of gravitational mass is borne out of the experimental observation that the external gravitational potential generated by a spherical body is proportional to $1/r$ \emph{independently of the composition and of the radius} of the body itself. It follows that to characterise completely the external gravitational f\mbox{}ield only one number is needed, the proportionality constant, from which the Newton's constant $G_{_{\!N}}$ is factored out for dimensional reasons. This number, the ``gravitational charge'' of the body, is conventionally called the gravitational mass. This notion is extended to the non-spherically symmetric case by selecting the monopole term in the multipole expansion of the external gravitational f\mbox{}ield, or equivalently considering the asymptotic behaviour of the latter (since the monopole term is the one with the slowliest decay).

General Relativity, although being more complex and having more degrees of freedom, shares the above mentioned property. Choosing the gauge suitably, the metric outside of a spherical body can always be written in the form
\begin{equation} \label{Schwarzschild sol}
	ds^2 = - \bigg( 1 - \frac{r_{\!\textup{g}}}{r} \bigg) dt^{2} + \bigg( 1 - \frac{r_{\!\textup{g}}}{r} \bigg)^{\!\!-1} dr^{2} + r^{2} \, \big( d\th^{2} + \sin^{2} \! \th \, d \vf^{2} \big) \quad ,
\end{equation}
so also in this case the external f\mbox{}ield is completely characterised by one number, the characteristic radius $r_{\!\textup{g}}\,$. Mathematically, the role of the source term (i.e.\ $\r$ and $p$, in our case) is to provide boundary conditions at the body's surface for the external solution (\ref{Schwarzschild sol}), conditions which are to be found by solving the equations of motion inside the body. The gravitational mass in GR is usually def\mbox{}ined by resorting to the Newtonian limit of the theory \cite{Misner 1973, Wald 1984}, that is setting 
\begin{equation} \label{GR mass def}
	M \equiv \frac{r_{\!\textup{g}}}{2G_{_{\!N}}} = \lim_{r \to \infty} \frac{r}{2 G_{_{\!N}}} \Big( 1 - B(r) \Big) = \lim_{r \to \infty} \frac{r}{2 G_{_{\!N}}} \bigg( 1 - \frac{1}{A(r)} \bigg) \quad .
\end{equation}
This def\mbox{}inition can be given a more elegant and coordinate-independent form, and generalised to stationary and asymptotically f\mbox{}lat space-times which are not spherically symmetric, such as in the Komar expression for $M$.\footnote{We remind that the Komar and ADM mass coincide for stationary, asymptotically f\mbox{}lat spacetimes if the initial data set of the ADM construction is chosen suitably \cite{Wald 1984}.} The rationale is that the gravitational mass is a measure of the external gravitational f\mbox{}ield produced by a body, it is a global quantity (in the sense that it is associated to the spacetime itself, not to the volume occupied by the star), and its value is linked to the ef\mbox{}fect the gravitational f\mbox{}ield has on test bodies in the Newtonian limit. The idea of identifying the physical parameters of isolated astrophysical bodies by studying the asymptotic behaviour of the gravitational f\mbox{}ield is indeed ubiquitous in gravitational physics.

One can however use the equations of motion to link the gravitational mass (def\mbox{}ined as above) to the properties of the source. Integrating in the Schwarzschild gauge (\ref{metric}) the time-time component of the Einstein equations, it is easy to derive the relation between $M$ and the energy density of the source
\begin{equation} \label{GR rho mass}
	M = 4 \pi \int_{0}^{r_{\star}} \!\! \r(r) \, r^{2} dr \quad .
\end{equation}
Since the energy density is a scalar, the expression on the right hand side of (\ref{GR rho mass}) is not invariant with respect to radial re-parametrisations. Indeed, including the proper volume element $\sqrt{g^{(3)}} dx^{3}$ on the constant-time spatial hypersurfaces, one obtains the \emph{proper mass} of the body \cite{Wald 1984}
\begin{equation} \label{proper mass}
	M_{\textup{p}} = \int_{V_{\star}} \r \, \sqrt{g^{(3)}} \, dx^{3} = 4 \pi \int_{0}^{r_{\star}} \! \r(r) \, \sqrt{A(r)} \, r^{2} dr \quad ,
\end{equation}
which is invariant ($V_{\star}$ indicates the volume occupied by the star). The fact that the equations of motion indicate (\ref{GR rho mass}) as the correct relation is interpreted, \emph{a posteriori}, assuming that the expression on the right hand side of (\ref{GR rho mass}) evaluated in Schwarzschild coordinates takes into account also the gravitational binding energy, thereby enforcing the equivalence between energy and gravitational mass for extended and self-gravitating objects.

It is customary to rephrase the discussion of the mass of spherical stars in terms of the so-called \emph{mass function}
\begin{equation} \label{mass function}
	\mcal{M}(r) = \frac{r}{2 G_{_{\! N}}} \, \bigg( 1 - \frac{1}{A(r)} \bigg) \quad .
\end{equation}
Indicating with $\mscr{B}_{r}(0)$ the sphere of radius $r$ centered at the origin, in Schwarzschild coordinates we have
\begin{equation}
	\int_{\mscr{B}_{r}(0)} G_{t}^{\,\,\, t} \, y^2 \, dy \, d\O = - 8 \pi G_{_{\! N}} \, \mcal{M}(r) \quad .
\end{equation}
The Einstein's equations imply that $\mcal{M}(r)$ is constant outside the star, where it coincides with the ef\mbox{}fective Newtonian mass (\ref{GR mass def}), and taking into account (\ref{stress energy st sph}) we get
\begin{equation} \label{mass function and rho}
	\mcal{M}(r) = 4 \pi \int_{0}^{r} \r(y) \, y^{2} dy \quad .
\end{equation}
The continuity of $\mcal{M}$ across the star's surface then implies (\ref{GR rho mass}). Although we cannot in general associate the concept of gravitational energy to a f\mbox{}inite volume, since there is no unique way to introduce a (local) gravitational stress-energy tensor, for the case of spherical stars (and just for this case) it is meaningful to interpret $\mcal{M}(r)$ as the gravitational mass enclosed in the sphere of radius $r$ \cite{Misner 1973}. This is a consequence of the non-existence of spherically symmetric gravitational radiation in Einstein's gravity. Considering non-static (but still spherical) conf\mbox{}igurations, it follows that the energy inside a sphere of radius $r$ can change only because of heat f\mbox{}luxes and radiation of particles across the sphere's surface, or by work done on the surface by pressure forces. Therefore, the fact that we can associate a (localised) mass $\mcal{M}(r)$ to the sphere of radius $r$ is due to the circumstance that in the spherical case any transfer of energy is detectable by local measurements.

Clearly this situation breaks down as long as we depart from spherical symmetry. On the other hand there is no dif\mbox{}f\mbox{}iculty in def\mbox{}ining the proper mass contained in any chosen spatial volume, independently of the conf\mbox{}iguration being spherically symmetric or not, since the proper mass does not include the gravitational binding energy.

\subsection{The case of \boldmath $R^{2}$-gravity}
\label{subsec: mass in Starobinsky gravity}

In $R^{2}$-gravity the situation is qualitatively dif\mbox{}ferent. As can be seen from the expressions (\ref{A sol ext}) and (\ref{B sol ext}) for the metric components in the weak-f\mbox{}ield approximation, the external gravitational f\mbox{}ield is determined by \emph{two} numbers, $r_{\!\z}$ and $r_{\!g}\,$, or in other words by two gravitational charges. This is a consequence of the fact that the metric components, beside by $\r$ and $p\,$, are sourced also by the scalar degree of freedom $\z$ which dynamically extends outside the body (being itself sourced by $\r$ and $p$ and obeying a Klein-Gordon equation). It follows that, for what concerns the metric, the energy density and the pressure do not generate simply a boundary condition for the external gravitational f\mbox{}ield. The impossibility of describing the external geometry with one gravitational charge, which is fundamentally due to the presence of the extra scalar DOF and therefore not peculiar to the weak-f\mbox{}ield approximation, has profound implications as we can see.

\subsubsection{On the usual def\mbox{}initions of gravitational mass}

It is quite easy to see that, in $R^{2}$-gravity, the relations (\ref{GR mass def}) and (\ref{GR rho mass}) are not compatible. Let us def\mbox{}ine for ease of notation
\begin{align} \label{asymptotic and rho masses}
	M_{\textup{g}} &= \lim_{r \to \infty} \, \frac{r}{2G_{_{\!N}}} \, \bigg( 1 - \frac{1}{A(r)} \bigg) & M_{\r} = 4 \pi \int_{0}^{r_{\star}} \!\! \r(r) \, r^2 dr \quad ,
\end{align}
and introduce the operator
\begin{equation}
	\mscr{D}_{\m}{}^{\n} = 2 \, \Big( g^{\nu\la} \, \nabla_{\!\mu} \, \de_{\la} - \d_{\m}{}^{\n} \, \Box \Big) \quad .
\end{equation}
Rewriting the equation (\ref{metric eq gen}) as follows
\begin{equation} \label{metric eq gen 2}
	G_{\m}{}^{\n} = \mscr{D}_{\m}{}^{\n} \, \z - 3 m^2 \, \z^2 \, \delta_{\m}{}^{\n} - 2 \z \, G_{\m}{}^{\n} + 8 \pi G_{_{\! N}} \, T_{\m}{}^{\n} \quad ,
\end{equation}
the integration of the $tt$ component on the sphere of radius $r$ gives
\begin{equation} \label{Mass function zeta and rho}
	\mcal{M}(r) = \frac{1}{8 \pi G_{_{\! N}}} \int_{\mscr{B}_{r}(0)} \bigg( \!- \mscr{D}_{t}{}^{t} \, \z + 3 m^2 \, \z^2 + 2 \z \, G_{t}{}^{t} \bigg) \, dV + 4 \pi \int_{0}^{r} \! \r(y) \, y^{2} dy \quad ,
\end{equation}
where $dV$ indicates the f\mbox{}lat volume element. Sending $r \to \infty$ we get
\begin{equation} \label{Mg - Mrho}
	M_{\textup{g}} - M_{\r} = \frac{1}{8 \pi G_{_{\! N}}} \int_{\mbbR^{3}} \bigg( \!- \mscr{D}_{t}{}^{t} \, \z + 3 m^2 \, \z^2 + 2 \z \, G_{t}{}^{t} \bigg) \, dV \quad .
\end{equation}
Note that, since $\z$ has a Yukawa behaviour with range $m^{-1}$, only the sphere of approximate radius $m^{-1} \sim \sqrt{\a}$ contributes to the integral. The quantity on the right hand side of (\ref{Mg - Mrho}) does not vanish in general, as can be checked numerically (see Section \ref{sec: numerical mass radius}), so the def\mbox{}initions (\ref{asymptotic and rho masses}) of gravitational mass are not equivalent in $R^{2}$-gravity.

It is important to understand which physical meaning can be given to $M_{\textup{g}}$ and $M_{\r}$ in this context, and whether one of the two def\mbox{}initions is eligible as ``the'' def\mbox{}inition of gravitational mass. In the asymptotic region the weak-f\mbox{}ield approximation is always valid, and therefore (\ref{A sol ext})-(\ref{B sol ext}) hold. Moreover, the terms containing $r_{\!\z}$ are exponentially suppressed. This means that $M_{\textup{g}}$ is the Newtonian mass felt by test-bodies orbiting in the asymptotic region, so $M_{\textup{g}}$ maintains the interpretation it has in GR although its spatial validity is limited (it is a ``distant observers'' mass). The story for $M_{\r}$ is more complicated. It may be thought that the failure of $M_{\r}$ to coincide with the asymptotic Newtonian mass $M_{\textup{g}}$ is due to $M_{\r}$ being linked to the properties of the spacetime near the star, while $M_{\textup{g}}$ is linked to the properties of the spacetime far away. This idea is however not correct, as can be seen as follows. Considering a conf\mbox{}iguration where $m r_{\star} \ll 1$ and the weak-f\mbox{}ield approximation holds, test-bodies orbiting close to the star (i.e.\ well inside the range of $\z$) feel an ef\mbox{}fective Newtonian mass equal to
\begin{equation} \label{nearby mass}
	M_{\textup{n}} = \frac{r_{\star}}{2G_{_{\! N}}} \, \Big( 1 - B(r_{\star}) \Big) = \frac{1}{2G_{_{\! N}}} \, \frac{r_{\!\z} + 3 r_{\!g}}{3} \quad ,
\end{equation}
and taking the limit $m \to 0$ we get
\begin{equation} \label{nearby mass m=0}
	M_{\textup{n}} = \frac{4}{3} M_{\r} - P \quad .
\end{equation}
Even neglecting the pressure we have $M_{\r} \neq M_{\textup{n}}$. Moreover, away from the limit $m \to 0$ the ratio between $M_{\r}$ and $M_{\textup{n}}$ does not remain constant (and equal to $4/3$ when $P = 0$), but displays a very complicated dependence on $m$, $r_{\star}$, $\r$ and $p\,$. Therefore $M_{\r}$ is not simply related to the ef\mbox{}fective Newtonian mass in any region of spacetime. This situation may be regarded as inconvenient enough to abandon the idea of interpreting $M_{\r}$ as a gravitational mass, as already pointed out in \cite{Resco:2016upv} where $M_{\r}$ is regarded as merely a parameter (a ``tag'') characterising families of solutions, without a specif\mbox{}ic physical interpretation.

\subsubsection{On a unique def\mbox{}inition of gravitational mass}

One may however take the totally opposite point of view, turning the impossibility of describing the external metric with a unique gravitational charge into an indication that in $R^{2}$-gravity the concept of gravitational mass has to be disentangled from the behaviour of test bodies in the Newtonian limit. This argument may be used in favour of regarding $M_{\r}$ as \emph{the} def\mbox{}inition of gravitational mass in $R^{2}$-gravity, owing to its interpretation as the total energy of matter.\footnote{We are here loosely using the word ``matter'' to mean ``matter, radiation and every form of energy-momentum included in $T_{\m}{}^{\n}\,$''.} Recall in fact that (in the Jordan frame) the stress-energy tensor couples only to the metric, in the sense that there is no direct coupling between $\z$ and $T\,$.
Moreover, the coupling term is exactly that of GR. This means in particular that matter feels only the spacetime metric or, in other words, that regarding the dynamical behaviour of matter the role of the scalar $\z$ is just to indirectly inf\mbox{}luence the metric conf\mbox{}iguration $g_{\m\n}$ via the equations of motion. Since it is assumed that the integral which def\mbox{}ines $M_{\r}$ correctly takes into account the gravitational binding energy in GR, it is reasonable to expect this be true also for theories where $T_{\m\n}$ couples only to $g_{\m\n}$ via the same coupling term.\footnote{Of course, a conf\mbox{}iguration $\big( \r(r), p(r) \big)$ cannot be an equilibrium prof\mbox{}ile at the same time for GR and $R^{2}$-gravity. But the binding energy is not def\mbox{}ined only for equilibrium conf\mbox{}igurations.}

Following this line of reasoning, the relation (\ref{Mg - Mrho}) would lend itself to the heuristic interpretation that the asymptotic mass $M_{\textup{g}}$ is given by the sum of the energy of matter (including the binding energy), i.e.\ $M_{\r}\,$, and the energy associated to the additional degree of freedom, i.e.\ the right hand side. This interpretation can be formalised by associating to the extra DOF $\z$ an ef\mbox{}fective stress-energy tensor $\mcal{T}_{\m}{}^{\n}$ def\mbox{}ined as follows
\begin{equation} \label{zeta stress energy 1}
	8 \pi G_{_{\! N}} \, \mcal{T}_{\m}{}^{\n} = \mscr{D}_{\m}{}^{\n} \, \z - 3 m^2 \, \z^2 \, \delta_{\m}{}^{\n} - 2 \z \, G_{\m}{}^{\n} \quad ,
\end{equation}
which is covariantly conserved as a consequence of the equation of motion (\ref{metric eq gen 2}). Accordingly, we may indicate $\mcal{R} = - \mcal{T}_{t}{}^{t}$ and consider it as the ef\mbox{}fective energy density of the extra DOF
\begin{equation} \label{zeta energy density}
	8 \pi G_{_{\! N}} \, \mcal{R} = - \mscr{D}_{t}{}^{t} \, \z + 3 m^2 \, \z^2 + 2 \z \, G_{t}{}^{t} \quad .
\end{equation}

It is worthwhile to point out that in the literature it is customary to introduce the ef\mbox{}fective stress-energy tensor associated to the extra DOF in a dif\mbox{}ferent way (see for example \cite{Capozziello:2002rd, Capozziello:2006uv}), to wit
\begin{equation} \label{zeta stress energy 2}
	8 \pi G_{_{\! N}} \, \mscr{T}_{\m}{}^{\n} = \frac{1}{1 + 2 \z} \, \bigg[ \, 2 \, \Big( g^{\nu\la} \, \nabla_{\!\mu} \, \de_{\la} - \d_{\mu}{}^{\nu} \, \Box \Big) \z - 3 m^2 \, \z^2 \, \delta_{\mu}{}^{\nu} \, \bigg] \quad ,
\end{equation}
so that the equation (\ref{metric eq gen}) becomes
\begin{equation} \label{cucu}
	G_{\m}{}^{\n} = 8 \pi G_{_{\! N}} \bigg( \mscr{T}_{\m}{}^{\n} + \frac{1}{1 + 2 \z} \, T_{\m}{}^{\n} \bigg) \quad .
\end{equation}
One may then integrate the $tt$ component of (\ref{cucu}) over the whole 3D space to derive a relation similar to (\ref{Mg - Mrho}), obtaining a dif\mbox{}ferent expression on the right hand side and, on the left hand side, the integral of $\r/(1 + 2 \z)$ instead of $M_{\r}\,$. It is not clear however what the interpretation of this relation should be. For example, it is doubtful that the integral of $\r/(1 + 2 \z)$ should be interpreted as the total matter energy. Moreover, both $\mscr{T}_{\m}{}^{\n}$ and $T_{\m}{}^{\n}/(1 + 2 \z)$ are not (separately) covariantly conserved (unless $\z$ is constant). For this reason, we believe (\ref{Mg - Mrho}) and (\ref{zeta stress energy 1}) to be physically more meaningful.

\subsubsection{Gravisphere and surface redshift}

According to the above interpretation of the relation (\ref{Mg - Mrho}), the dif\mbox{}ference between $M_{\textup{g}}$ and $M_{\r}$ is to be found in the contribution of the extra DOF, both inside and outside the star. This may suggest that a useful characterisation of the properties of spacetime outside and near the star may be given by introducing a new def\mbox{}inition of gravitational mass which, if possible, takes into account both the contribution of matter and of the extra DOF inside the star's surface.

Such a def\mbox{}inition has indeed been proposed in \cite{Astashenok:2017dpo}, by suitably using the function $\mcal{M}(r)$ def\mbox{}ined in (\ref{mass function}) (see also \cite{Resco:2016upv}). More specif\mbox{}ically, the attention is cast upon the quantity
\begin{equation} \label{stellar mass}
	M_{\textup{s}} = \mcal{M}(r_{\!\star}) = \frac{r_{\!\star}}{2 G_{_{\! N}}} \, \bigg( 1 - \frac{1}{A(r_{\!\star})} \bigg) \quad ,
\end{equation}
which is referred to as the ``stellar mass bounded by the star's surface'', as opposed to the mass $M_{\textup{g}} = \lim_{r \to \infty} \mcal{M}(r)$ which is referred to as the ``gravitational mass measured by a distant observer''. Evaluating the relation (\ref{Mass function zeta and rho}) at $r = r_{\star}$ we obtain
\begin{equation} \label{Ms and Mrho}
	M_{\textup{s}} = M_{\r} + \int_{V_{\star}} \mcal{R} \, dV \quad ,
\end{equation}
which, again assuming that $\mcal{T}_{\m}{}^{\n}$ indeed is the legitimate stress-energy tensor for the extra DOF, indeed suggests that $M_{\textup{s}}$ takes into account also the energy of the extra DOF inside the star. Here $dV$ indicates the f\mbox{}lat space volume element. Indicating with $V_{\star}^{^{C}}$ the (set) complement of $V_{\star}\,$, or in other words the region outside the star, the relation (\ref{Mg - Mrho}) can then be rewritten as
\begin{equation} \label{Mg and Ms}
	M_{\textup{g}} = M_{\textup{s}} + \int_{V_{\star}^{^{C}}} \mcal{R} \, dV \quad ,
\end{equation}
where, as for (\ref{Mg - Mrho}), only the spherical shell of radii $r_{\!\star} < r \lesssim m^{-1}$ signif\mbox{}icantly contributes to the integral. We may then interpret (\ref{Mg and Ms}) as if the dif\mbox{}ference between $M_{\textup{g}}$ and $M_{\textup{s}}$ were given by the energy associated to the extra DOF in the above mentioned spherical shell. The analysis of \cite{Astashenok:2017dpo} is indeed centered on the idea that the curvature present inside a region called gravisphere, which surrounds the star and has external radius approximately equal to $\sqrt{\a}\,$, itself contributes to the the distant observer's mass.

Clearly, assigning a mass to the volume enclosed by the star's surface (and, complementary, to the gravisphere) raises immediate concerns, since (\ref{stellar mass}) is meant to include the gravitational binding energy (it is not a proper mass in the sense of (\ref{proper mass})), which in general cannot be localised. Remember in fact that in GR the identif\mbox{}ication of $\mcal{M}(r)$ with the mass contained inside the sphere of radius $r$ is possible only thanks to the absence of spherically symmetric gravitational waves; however, spherically symmetric radiation do exists in $R^{2}$-gravity, being the (scalar) waves of $\z\,$. The validity of the interpretation of \cite{Astashenok:2017dpo} therefore strongly relies on our ability of def\mbox{}ining a local and covariantly conserved stress-energy tensor for $\z\,$, the obvious candidate being (\ref{zeta stress energy 1}). If, in the spherically symmetric and time-dependent case, one trusts (\ref{zeta stress energy 1}) to describe the f\mbox{}lux of energy and momentum associated to $\z$ across the star's surface, then also in this case $M_{\textup{s}}$ can change only because of f\mbox{}luxes of energy and momentum detectable by local measurements (now including also scalar radiation).

The mass $M_{\textup{s}}$ is without doubt a legitimate ``tag'' to characterise the spherically symmetric solutions. It is however important to point out that, contrary to the claim of \cite{Astashenok:2017dpo}, $M_{\textup{s}}$ does not characterise to the surface gravitational redshift of the star \cite{Van Paradijs 1979}. The gravitational redshift $z_{\star}$ undergone by an electromagnetic wave emitted (with frequency $\o_{e}$) on the surface of the star and detected (with frequency $\o_{d}$) in the asymptotic region $r \gg m^{-1}$ is given by
\begin{equation}
	z_{\star} = \frac{\o_{e} - \o_{d}}{\o_{d}} = \frac{1}{\sqrt{\abs{g_{tt}(r_{\!\star})}\,}} - 1 = \frac{1}{\sqrt{B(r_{\!\star})\,}} - 1 \quad ,
\end{equation}
so it is controlled by the $tt$ component of the metric on the surface. On the other hand $M_{\textup{s}}$ is determined by the $rr$ component of the metric on the star's surface, and in $R^2$-gravity there is no simple and general relation between $\abs{g_{tt}(r_{\!\star})}$ and $g_{rr}(r_{\!\star})$ (unlike in GR, where they are one the inverse of the other). We could def\mbox{}ine an ef\mbox{}fective ``surface redshift'' mass $M_{\textup{sr}}$ by means of the relation
\begin{equation}
	z_{\star} = \frac{1}{\sqrt{1 - \dfrac{2 G_{_{\! N}} M_{\textup{sr}}}{r_{\!\star}} \rule{0pt}{6mm}}} - 1
\end{equation}
which holds in GR between $z_{\star}$ and the mass, that is def\mbox{}ining
\begin{equation}
	M_{\textup{sr}} = \frac{r_{\!\star}}{2 G_{_{\! N}}} \, \frac{z_{\star} \, (2 + z_{\star})}{\big( 1 + z_{\star} \big)^{2}} \quad .
\end{equation}
The ef\mbox{}fective mass $M_{\textup{sr}}$ def\mbox{}ined this way is dif\mbox{}ferent from $M_{\textup{s}}\,$, and indeed is none else than the nearby mass $M_{\textup{n}}\,$, introduced in (\ref{nearby mass}). The latter seems therefore better suited than $M_{\textup{s}}$ to describe the gravitational f\mbox{}ield in the proximity of the star.

\section{Neutron stars and numerical mass-radius relations}
\label{sec: numerical mass radius}

To render our analysis more concrete we now turn to the case of neutron stars in $R^{2}$-gravity and their M--R relation. This gives us the possibility of assessing quantitatively the relevance of the dif\mbox{}ference between the several def\mbox{}initions of mass described above, in a context where these dif\mbox{}ferences may be potentially relevant. See \cite{Astashenok:2013vza, Astashenok:2014pua, Astashenok:2014dja, Nari:2018aqs} for related work.

To achieve this, we numerically solve the equations (\ref{zeta eq}) (\ref{Aeqfull}) and (\ref{Beqfull}) by means of a shooting method. For def\mbox{}initeness we model the interior of the neutron star with the Sly equation of state \cite{Chabanat:1997un, Douchin:2001sv} expressed by the following analytic formula \cite{Haensel:2004nu}:
\begin{multline}
	\log_{10}p = \frac{a_1 + a_2\log_{10}\rho + a_3(\log_{10}\rho)^3}{\exp[a_5(\log_{10}\rho - a_6)] + 1}\frac{1}{1 + a_4\log_{10}\rho} + \frac{a_7 + a_8\log_{10}\rho}{\exp[a_9(a_{10} - \log_{10}\rho)] + 1} + \\[2mm]
	+ \frac{a_{11} + a_{12}\log_{10}\rho}{\exp[a_{13}(a_{14} - \log_{10}\rho)] + 1} + \frac{a_{15} + a_{16}\log_{10}\rho}{\exp[a_{17}(a_{18} - \log_{10}\rho)] + 1} \quad ,
\end{multline}
where the pressure is given here in units of dyn/cm$^2$ and the density in units of g/cm$^3$, and the 18 coef\mbox{}f\mbox{}icients $a_i$ come from a numerical f\mbox{}it and are tabulated in \cite{Haensel:2004nu}. As done in \cite{Sbisa:2018rem}, prior to the numerical integration we recast the system of equations in terms of dimensionless quantities. For example, we normalise the parameter $\a$ to the Sun's half Schwarzschild radius $r_{_{\!0}} = G_{_{\! N}} M_{_{\odot}}/c^2 \simeq 1.5 \; \mbox{km}$, so we work with the adimensional parameter $\hat{\a} = \a/r_{_{\!0}}^2$. Regarding the shooting, whose initial conditions are given at the centre of the star, the only free parameters are the central density $\r_{_{0}}$ and the central value $\z_{_{0}}$ of the extra DOF ($\sim$ curvature), since the central pressure is determined by $\r_{_{0}}$ via the equations of state. To individuate the solution we choose a priori a grid of values for the central density, and for each of them we determine $\z_{_{0}}$ via the shooting method, selecting the solution for $\z$ which decays exponentially to zero far from the star (meaning at $r \gg \sqrt{\a} \sim m^{-1}$). It follows that our shooting solutions, and therefore our curves in the M--R plane, are parametrised by the value of the central density. The star's radius is determined to be that for which $\r$ vanishes (strictly speaking, for which it becomes negative). We refer to \cite{Sbisa:2018rem} for a more detailed discussion of our procedure.\footnote{Again we remind that in \cite{Sbisa:2018rem} slightly dif\mbox{}ferent conventions are used, in that the role of $A$ and $B$ is interchanged.}

For def\mbox{}initeness we consider the representative values $\hat{\a} = 1$, $10$ and $100\,$, which permit to gain an understanding of how the properties of the star depend on the parameter $\a\,$.

\subsection{Numerical results}

\subsubsection{The asymptotic mass $M_{\textup{g}}$}

\begin{figure}[hbtp]
\includegraphics[width=0.55\columnwidth]{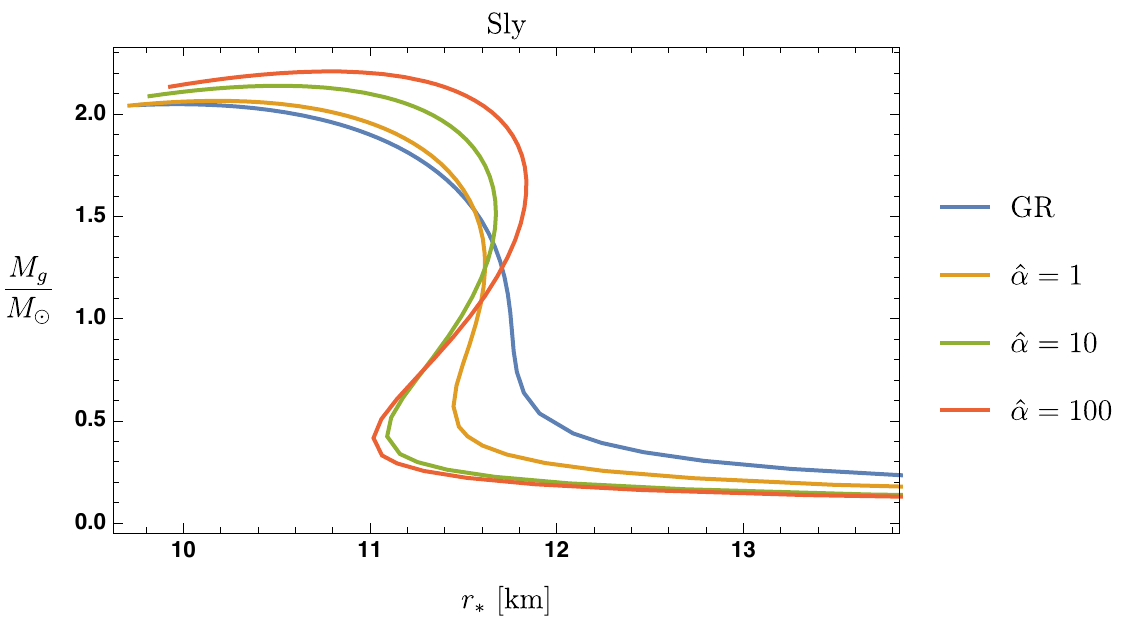} \includegraphics[width=0.55\columnwidth]{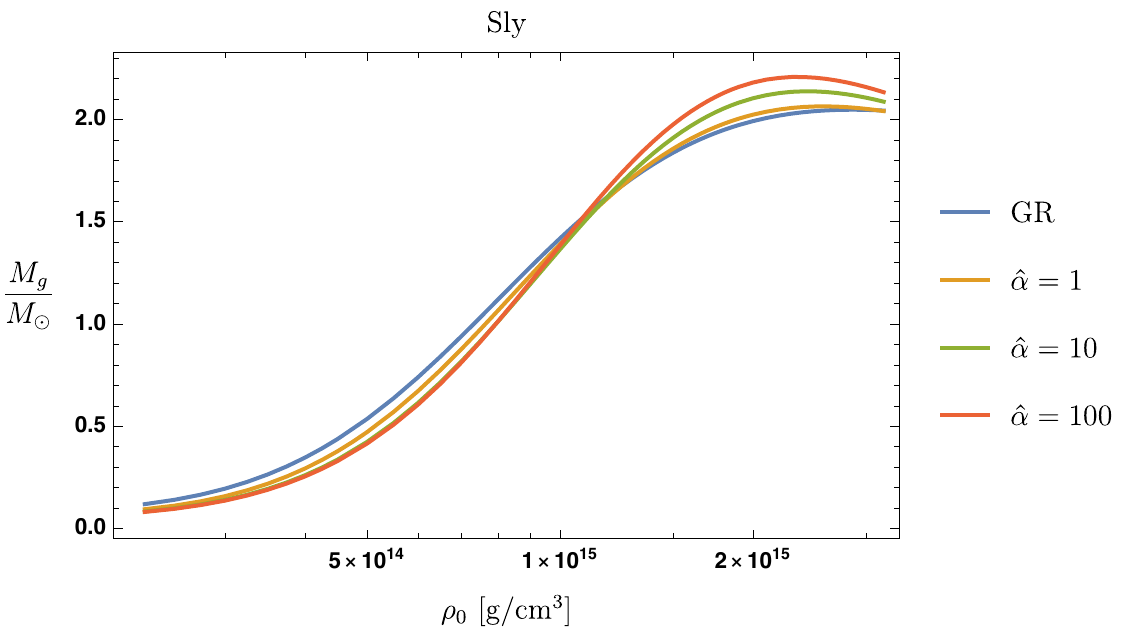}
\caption{The asymptotic mass $M_{\textup{g}}$ as function of the radius (left) and of the central density (right) for $\hat{\a} = 1, 10, 100\,$.} \label{Figure: Manhattan}
\end{figure}

Let us begin our discussion by considering the results for the asymptotic mass $M_{\textup{g}}\,$. Doing so allows to clearly highlight the main features of the curves and their dependence on $\hat{\a}\,$, and facilitates the subsequent comparison with the other def\mbox{}initions of mass. In Fig.~\ref{Figure: Manhattan} the M--R relation (left) and the dependence of $M_{\textup{g}}$ with the central density (right) are shown for the three chosen values of $\hat{\a}\,$.
These results are in qualitative agreement with the f\mbox{}indings of \cite{Yazadjiev:2014cza, Staykov:2014mwa, Yazadjiev:2015zia, Yazadjiev:2015xsj}. The central density increases monotonically along the M--R curves, the lowest values of $\r_{_{0}}$ corresponding to the bottom-right part of the curves and the highest values to top-left part. In particular, the mass increases monotonically with $\r_{_{0}}\,$. It is apparent that the maximum mass can get comfortably well above two solar masses, and that it increases with increasing $\a\,$.

An evident feature of the M--R curves is the presence of an intermediate mass range where the radius \emph{increases} with increasing mass and central density. In GR this behaviour is usually associated with a thermodynamic instability, which sets in at the points where the derivative $d M/d r_{\!\star}$ vanishes (turning point instability \cite{Sorkin:1981jc}, for a review see \cite{Green:2013ica}).\footnote{At least for most equations of state, Sly included. There are indeed cases, such as self-bound strange stars and stars with condensates, where the radius can increase with the total mass even in GR. \cite{Ozel:2016oaf}} This does not immediately imply an instability in $R^2$-gravity, since the $d M/d r_{\!\star} > 0$ part of the M--R curves is delimited by points where $d M/d r_{\!\star}$ diverges (or equivalently, where $d r_{\!\star}/d M$ vanishes). A hint to the presence of two regimes is visible also in the M-$\r_{_{0}}$ curves, where the value $M_{\textup{g}} \sim 1.5 M_{_{\odot}}$ separates two dif\mbox{}ferent behaviours. For smaller values of $M_{\textup{g}}$, the neutron star's mass in $R^2$-gravity is smaller (and even more so the higher $\a$ becomes) than the analogous star in GR with the same central density. The opposite behaviour happens for values of $M_{\textup{g}} \gtrsim 1.5 M_{_{\odot}}\,$. 

\subsubsection{The def\mbox{}initions of mass}
\label{subsubsec: definitions of mass}

\begin{figure}[hbtp]
	\includegraphics[width=0.55\columnwidth]{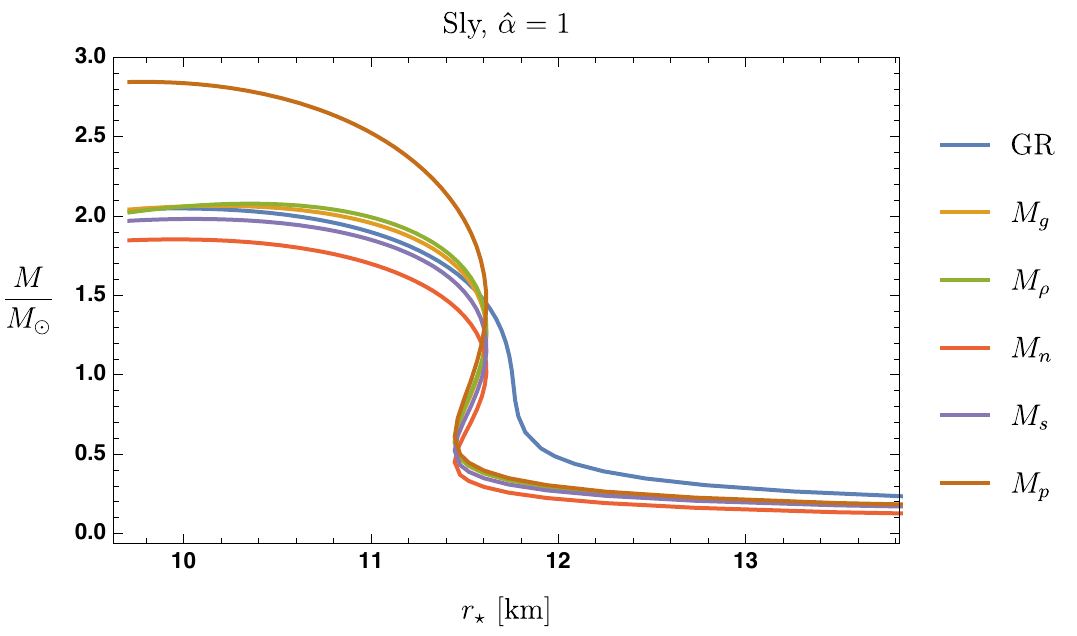} \includegraphics[width=0.55\columnwidth]{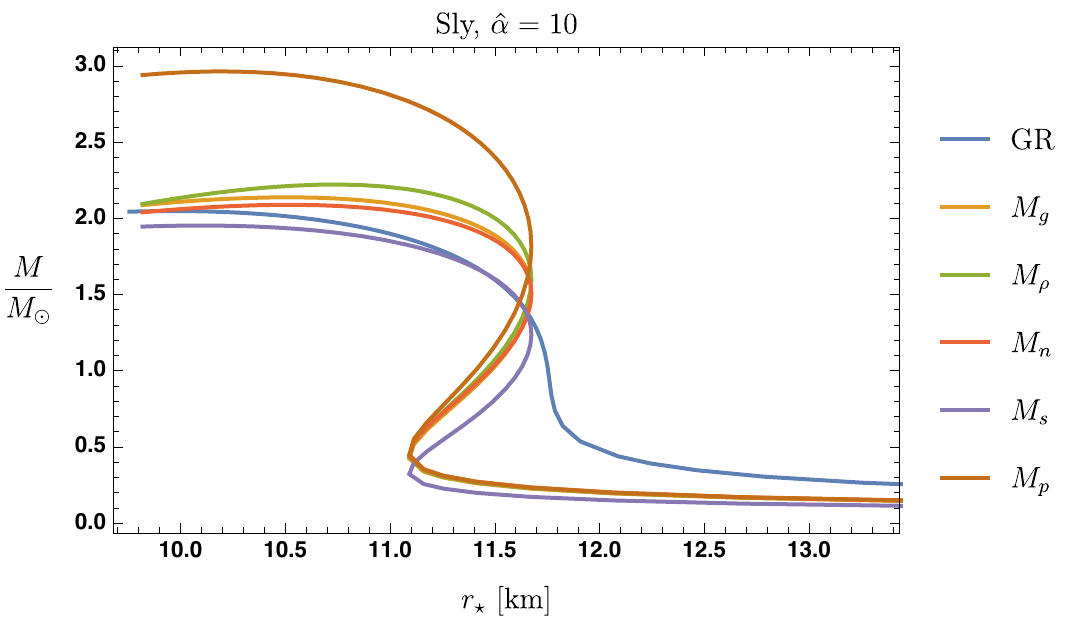} \includegraphics[width=0.55\columnwidth]{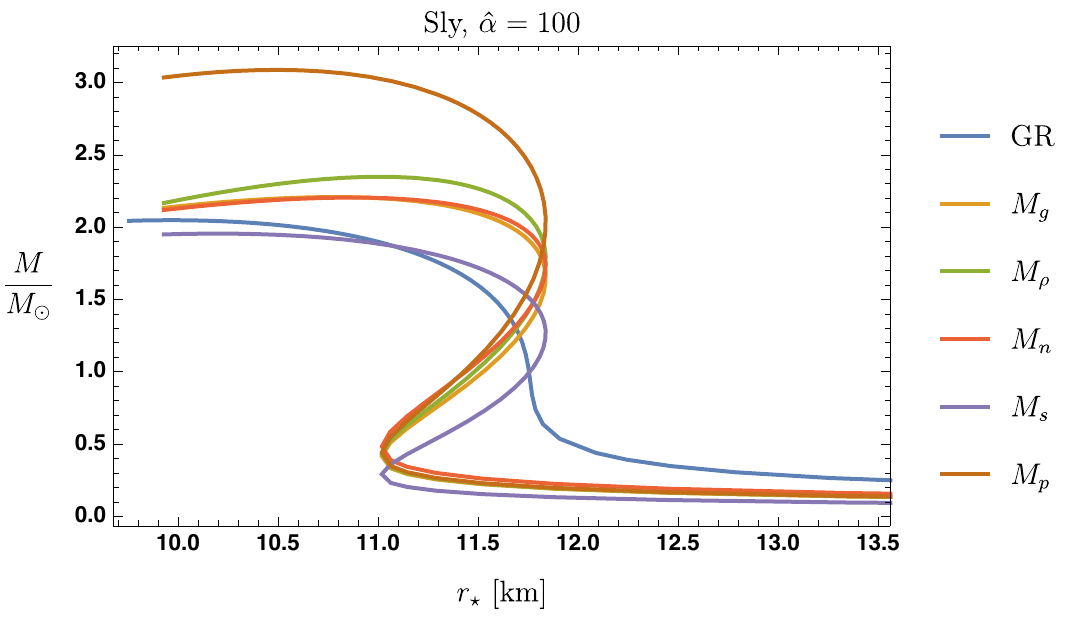}
	\caption{The M--R curves relative to the dif\mbox{}ferent def\mbox{}initions of mass, respectively for $\hat{\a} = 1$, $\hat{\a} = 10$ and $\hat{\a} = 100\,$. The GR curve is present in all the plots.} \label{Figure: Brazilian}
\end{figure}

Let us now compare the M--R curves relative to the def\mbox{}initions of mass discussed in Section \ref{sec: definition of mass}. In Figure \ref{Figure: Brazilian}, the M--R curves relative to $M_{\textup{g}}$, $M_{\r}$, $M_{\textup{p}}$, $M_{\textup{n}}$ and $M_{\textup{s}}$ are displayed respectively for $\hat{\a} = 1$, $\hat{\a} = 10$ and $\hat{\a} = 100\,$, with the curve for GR ($\hat{\a} = 0$) included in all the plots for the sake of comparison.

Several comments are in order. On general grounds note that, despite the dif\mbox{}ference in the curves, all of them share the feature mentioned above of having a region where $dr_{\!\star}/dM > 0$, which is the more evident the bigger the value of $\a\,$. Moreover, all the curves nearly coincide for small central density, while the dif\mbox{}ference is more pronounced when we approach the maximum mass. Note that there is a perceptible dif\mbox{}ference between the small mass limit of $R^2$-gravity and that of GR: this is not surprising, since the weak f\mbox{}ield limit of $R^2$-gravity is dif\mbox{}ferent from GR's. This dif\mbox{}ference becomes less important when $\a$ gets smaller, coherently with GR being reproduced in the $\a \to 0$ limit of $R^2$-gravity.

\begin{figure}[hbtp]
	\includegraphics[width=0.5\columnwidth]{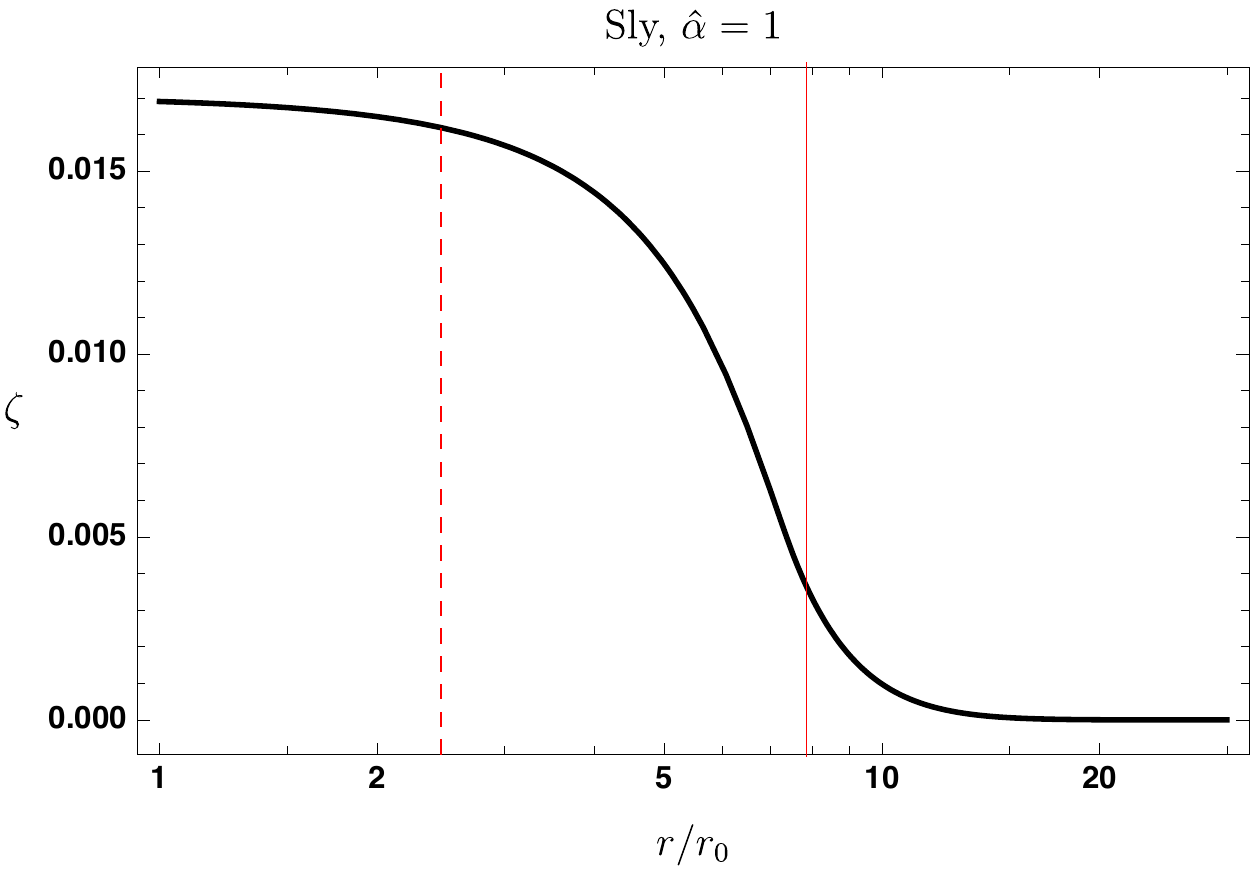} \includegraphics[width=0.5\columnwidth]{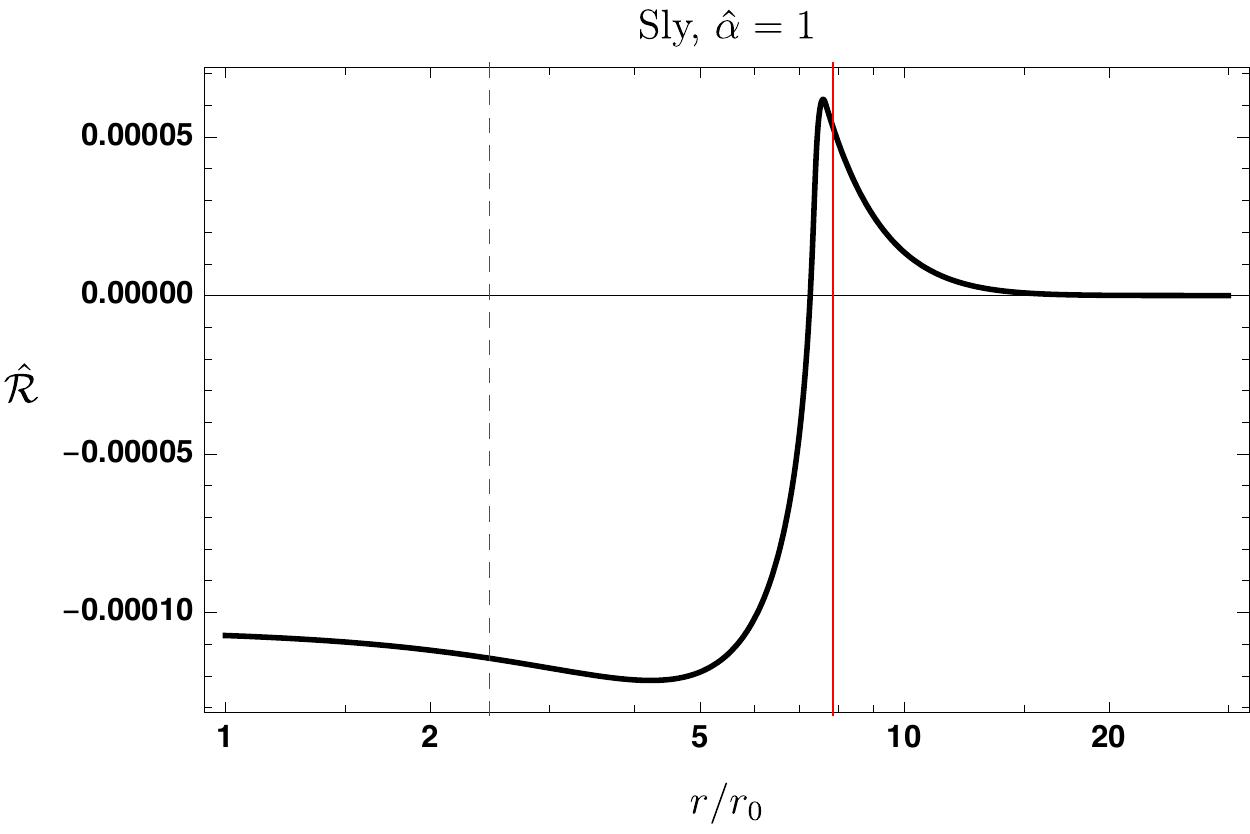}
	\caption{The extra DOF $\z$ (left) and the adimensional quantity $\hat{\mcal{R}}$ (right), as functions of $r/r_{_{0}}\,$. The continuous vertical line marks the star's surface while the dashed line marks the range of $\z\,$.} \label{Figure: Haitian}
\end{figure}

Focusing on the region close to the maximum mass, it is apparent that the proper mass is signif\mbox{}icantly higher than the others. This is sensible since the gravitational binding energy is negative, and we are considering compact objects where gravity is not weak. It is also apparent that $M_{\textup{s}}$ is smaller than $M_{\textup{g}}$ and $M_{\r}\,$, and signif\mbox{}icantly so when $\hat{\a} = 10$ and $\hat{\a} = 100\,$. This can be understood, in terms of our analysis of Section \ref{sec: definition of mass}, by referring to the behaviour of the extra DOF. We found that, generically, the function $\z(r)$ is concave inside the star and convex outside. Note that the ef\mbox{}fective energy density $\mcal{R}$ explicitly reads
\begin{equation} \label{zeta energy density explicit}
	8 \pi G_{_{\! N}} \, \mcal{R} = \frac{2}{A} \, \bigg[ \, \z^{\p\p} + \bigg( \frac{2}{r} - \frac{A^{\p}}{2 A} \bigg) \z^{\p} \, \bigg] + 3 m^2 \, \z^2 + 2 \z \, G_{t}{}^{t} \quad ,
\end{equation}
so its sign is inf\mbox{}luenced by the second derivative of $\z\,$. If $\z^{\p\p}$ dominates the other terms inside the star, then $\mcal{R}$ is negative inside the star and positive outside, so the contribution of $\mcal{R}$ is negative in Eq.\ (\ref{Ms and Mrho}) and positive in Eq.\ (\ref{Mg and Ms}), explaining why $M_{\textup{s}}$ is both smaller than $M_{\textup{g}}$ and $M_{\r}\,$. Furthermore, since the Equation \eqref{Mg - Mrho} can be rewritten as
\begin{equation} \label{Mg - Mrho new}
	M_{\textup{g}} - M_{\r} = \int_{\mbbR^{3}} \mcal{R} \, dV \quad ,
\end{equation}
the net ef\mbox{}fect of the competition between the negative (inside) and the positive (outside) contributions of $\mcal{R} \,$, with the preponderance of the former, is a relatively small dif\mbox{}ference between $M_{\textup{g}}$ and $M_{\r}\,$. This qualitative analysis is corroborated by the numerical results for $\z$ and for the adimensional quantity $\hat{\mcal{R}} = \mcal{R} \, G_{_{\! N}} r_{_{0}}^{2}/c^{4}$ as functions of the adimensional radius $r/r_{_{0}}\,$, plotted in Figure \ref{Figure: Haitian} for the case $\hat{\a} = 1$, $\r_{_{0}} = 10^{15} \,\, \text{g}/\text{cm}^{3}$ e $\z_{_{0}} = 0.0170299 \,$.

\subsubsection{Numerical values}
\label{subsubsec: Numerical values}

For completeness, we provide here a quantitative comparison between the M--R curves relative to dif\mbox{}ferent def\mbox{}initions of mass. To do this, we concentrate on the characteristic features of the curves mentioned above, to wit the maximum mass $M^{\textup{max}}$ and the mass range $\D M$ where $dr_{\!\star}/dM$ is positive. In Table \ref{tab: M max Sly} and \ref{tab: delta M Sly} we list the values obtained numerically for these quantities, for all the def\mbox{}initions of mass discussed above. Since our datapoints are discrete, we evaluate $M^{\textup{max}}$ by approximating the maximum point of the M--R curve with the maximum-value datapoint. We give the numerical value with three signif\mbox{}icant digits, without attempting to assess the uncertainty. Regarding $\D M$, we estimate the minimum and maximum mass of the interval by using the mass of the datapoint where the radius has an extremum (again, extremum with respect to the other datapoints). The error is estimated to be the dif\mbox{}ference in mass with the datapoints adjacent to the extremum.
\begin{table}[htb!]
\centering
\[
\begin{array}{cccccc}
\toprule
\hat{\a} \quad & \quad M^{\textup{max}}_{\textup{g}} \quad & \quad M^{\textup{max}}_{\r} \quad & \quad M^{\textup{max}}_{\textup{p}} \quad & \quad M^{\textup{max}}_{\textup{s}} \quad & \quad M^{\textup{max}}_{\textup{n}} \quad \\
\midrule
\textsc{gr} & 2.05 & 2.05 & 2.84 & 2.05 & 2.05 \\
1 & 2.06 & 2.08 & 2.85 & 1.98 & 1.85 \\
10 & 2.14 & 2.22 & 2.96 & 1.95 & 2.09 \\
100 & 2.21 & 2.35 & 3.09 & 1.96 & 2.20 \\
\bottomrule
\end{array}
\]
\caption{Maximum masses for GR and for $\hat{\a} = 1$, $10$, $100$. Solar mass units are employed.}
\label{tab: M max Sly}
\end{table}

\begin{table}[htb!]
\centering
\[
\begin{array}{cccccc}
\toprule
\hat{\a} & \D M_{\textup{g}} & \D M_{\r} & \D M_{\textup{p}} & \D M_{\textup{s}} & \D M_{\textup{n}} \\
\midrule
1 \quad & \quad 0.6 - 1.2 \quad & \quad 0.6 - 1.3 \quad & \quad 0.6 - 1.4 \quad & \quad 0.5 - 1.1 \quad & \quad 0.4 - 1.0 \\
10 \quad & \quad 0.4 - 1.5 \quad & \quad 0.4 - 1.6 \quad & \quad 0.4 - 1.8 \quad & \quad 0.3 - 1.2 \quad & \quad 0.4 - 1.5 \\
100 \quad & \quad 0.4 - 1.7 \quad & \quad 0.4 - 1.8 \quad & \quad 0.4 - 2.0 \quad & \quad 0.3 - 1.3 \quad & \quad 0.5 - 1.7 \\
\bottomrule
\end{array}
\]
\caption{Mass interval where $dr_{\!\star}/dM > 0\,$. Solar mass units are employed, and the error is $\pm 0.1\,$.}
\label{tab: delta M Sly}
\end{table}

It is apparent that, when $\hat{\a}$ lies in the range between $1$ and $100$, the maximum masses $M^{\textup{max}}_{\textup{g}}$, $M^{\textup{max}}_{\r}$ and $M^{\textup{max}}_{\textup{p}}$ roughly display a logarithmic behaviour $M^{\textup{max}}(\hat{\a}) = M^{\textup{max}}(1) + k \, \log_{10} (\hat{\a})\,$, with $k \simeq 0.07$, $\simeq 0.13$ and $\simeq 0.12$ respectively. We are not sure whether or not this approximate behaviour continues to hold for larger values of $\hat{\a}\,$. The behaviour of $M^{\textup{max}}_{\textup{s}}$ and $M^{\textup{max}}_{\textup{n}}$ are instead peculiar. In particular, the latter is signif\mbox{}icantly lower than the other masses when $\hat{\a} = 1$ but becomes very close to $M^{\textup{max}}_{\textup{g}}$ when $\hat{\a} = 100\,$. It is worthwhile to recall that $M^{\textup{max}}_{\textup{n}}$ can be interpreted as the mass felt by orbiting test bodies only for $\hat{\a}$ suf\mbox{}f\mbox{}iciently big, i.e.\ when the nearby region exists. For example this does not happen when $\hat{\a} = 1\,$, since in that case the range of the extra DOF is smaller than the radius of the star.

Regarding the mass range $\D M$, it is interesting to note that quite generically the interval becomes wider as $\hat{\a}$ increases, but without shifting signif\mbox{}icantly its position. In particular, the lower extreme of the interval decreases between $\a = 1$ and $\a = 10$ and then remains approximately constant between $\a = 10$ and $\a = 100\,$. Again, the case of the nearby mass is peculiar.

\subsection{Equation of state and degeneracy}

An important question is how (and how much) the results of the previous section depend on the equation of state. Specif\mbox{}ically, we would like to understand whether there is degeneracy between the EoS and the prof\mbox{}iles of the M--R curves relative to the various def\mbox{}initions of gravitational mass. While a thorough investigation is beyond the scope of the present work, it is worthwhile to perform here a preliminary investigation to shed light on this point. Therefore, we consider below the equations of state BSk19 and BSk20 \cite{Potekhin:2013qqa}, and perform again the analysis done above for the Sly EoS.

\subsubsection{Qualitative comparison}

\begin{figure}[hbtp]
	\includegraphics[width=0.55\columnwidth]{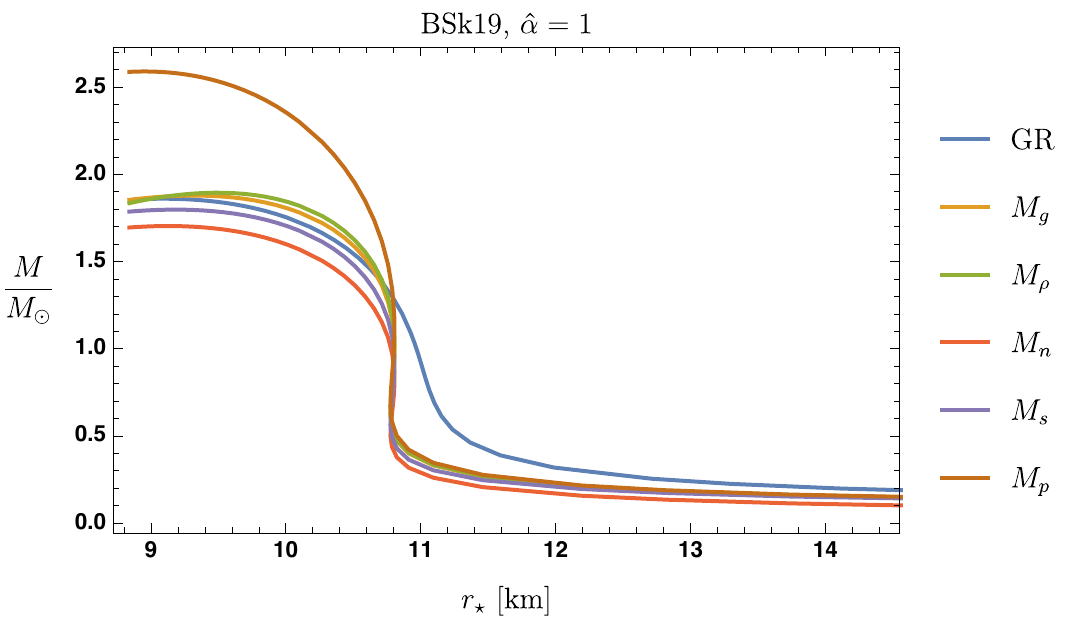} \includegraphics[width=0.55\columnwidth]{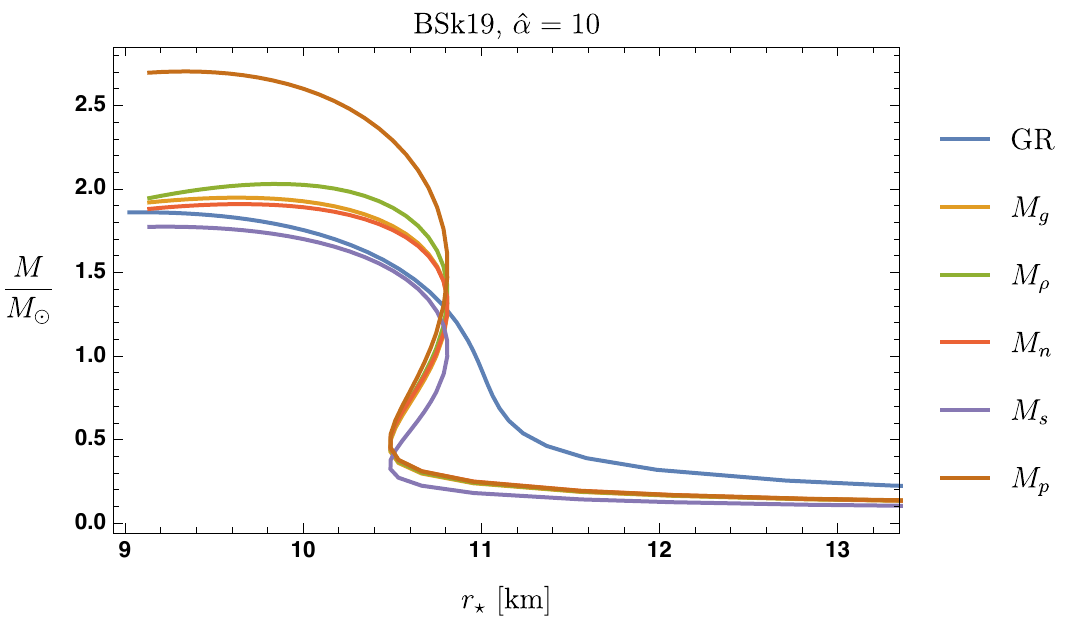} \includegraphics[width=0.55\columnwidth]{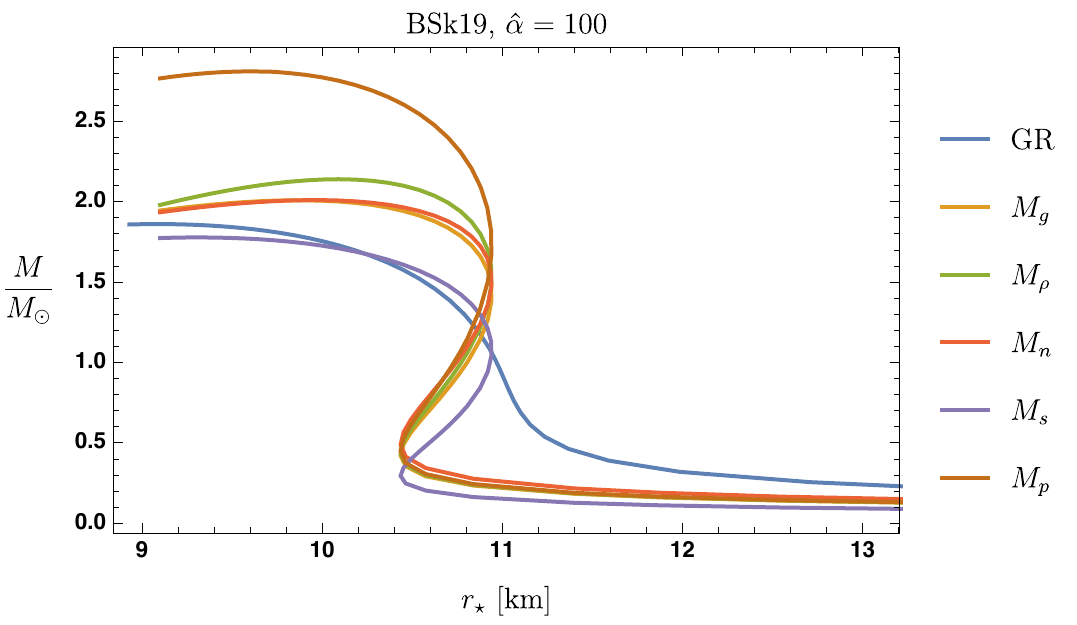} \includegraphics[width=0.55\columnwidth]{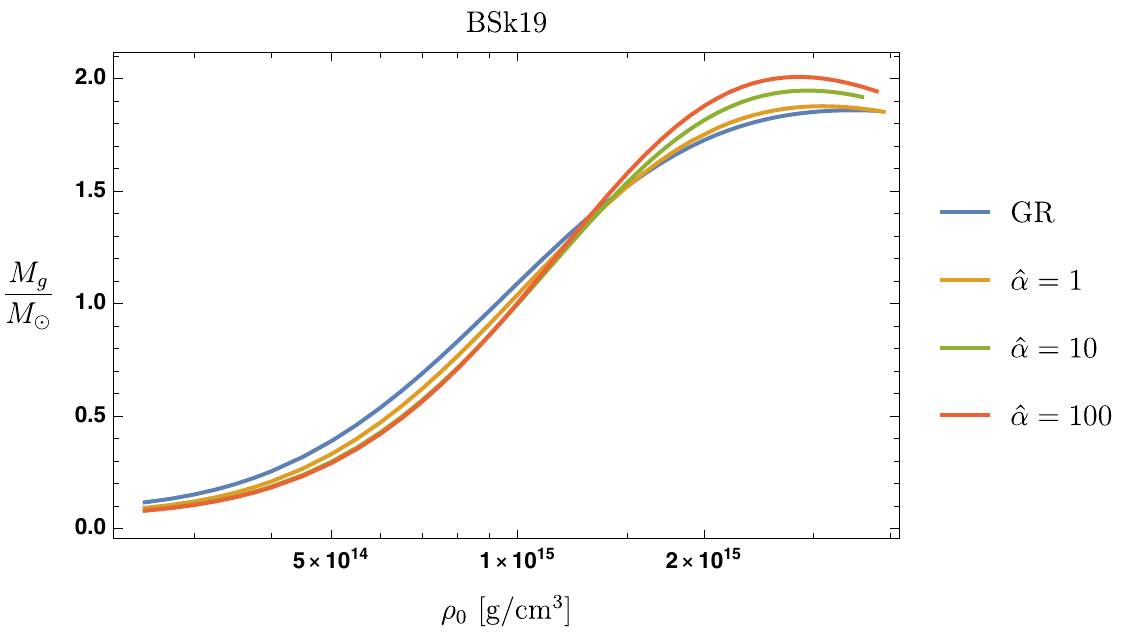}
	\caption{The M--R curves relative to the dif\mbox{}ferent def\mbox{}initions of mass for the BSk19 EoS, respectively for $\hat{\a} = 1$, $\hat{\a} = 10$ and $\hat{\a} = 100\,$. The GR curve is present in all the plots. The bottom-right plot shows the M--$\r_{_{0}}$ curves for the asymptotic mass $M_{\textup{g}}$.} \label{Figure: BSk19 curves}
\end{figure}

\begin{figure}[hbtp]
	\includegraphics[width=0.55\columnwidth]{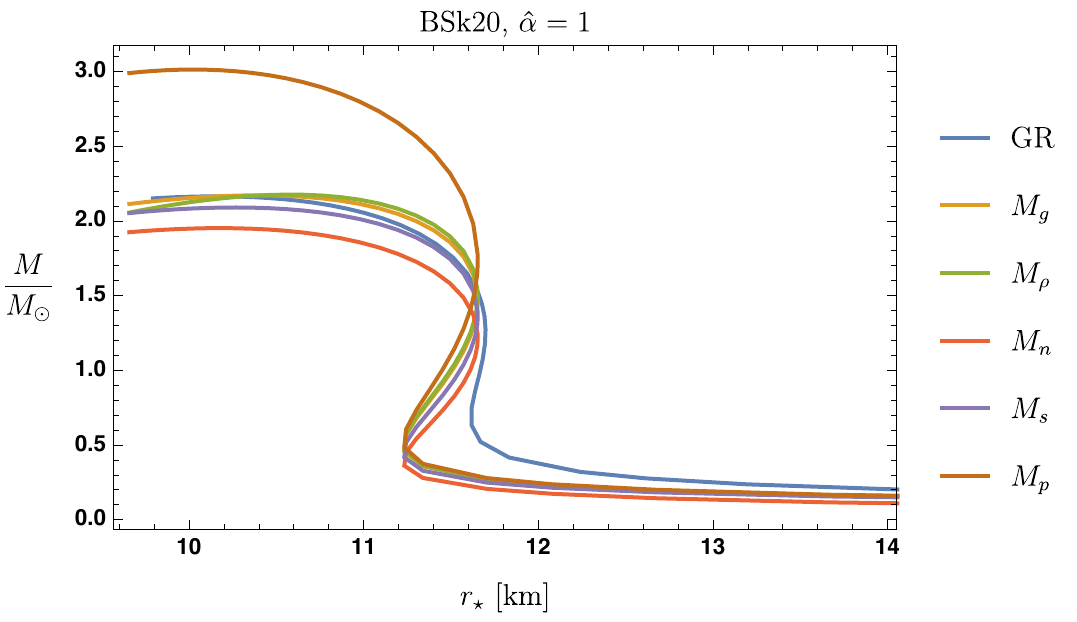} \includegraphics[width=0.55\columnwidth]{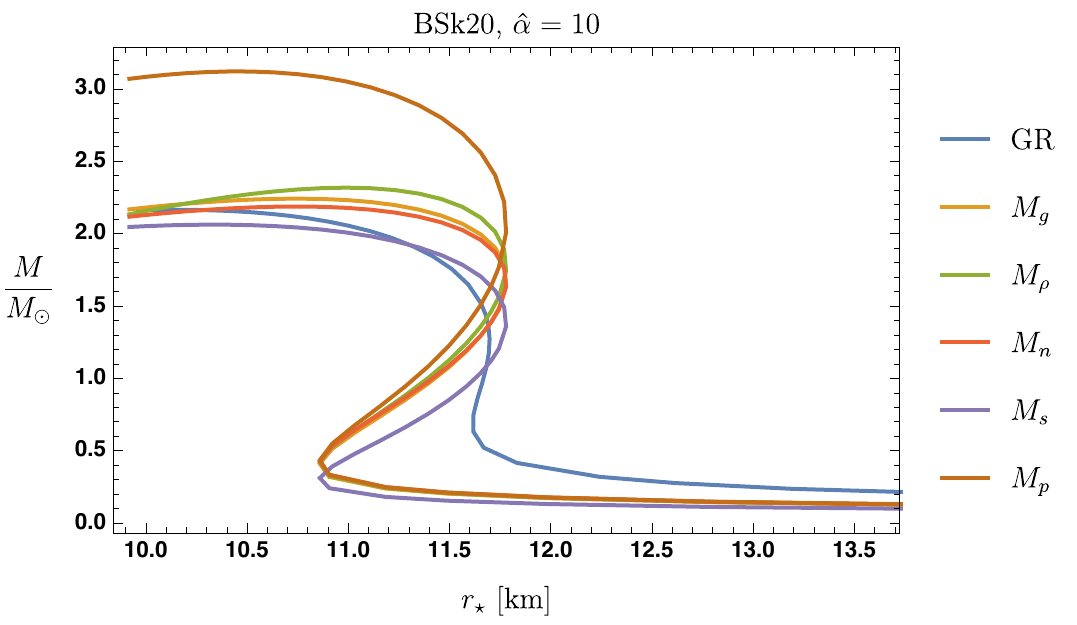} \includegraphics[width=0.55\columnwidth]{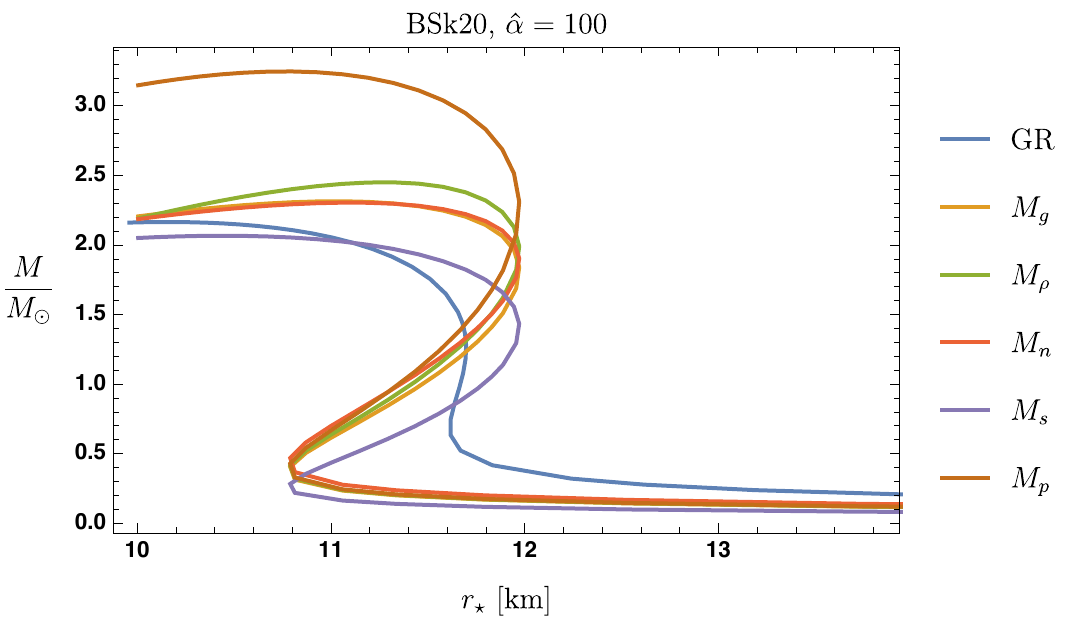} \includegraphics[width=0.55\columnwidth]{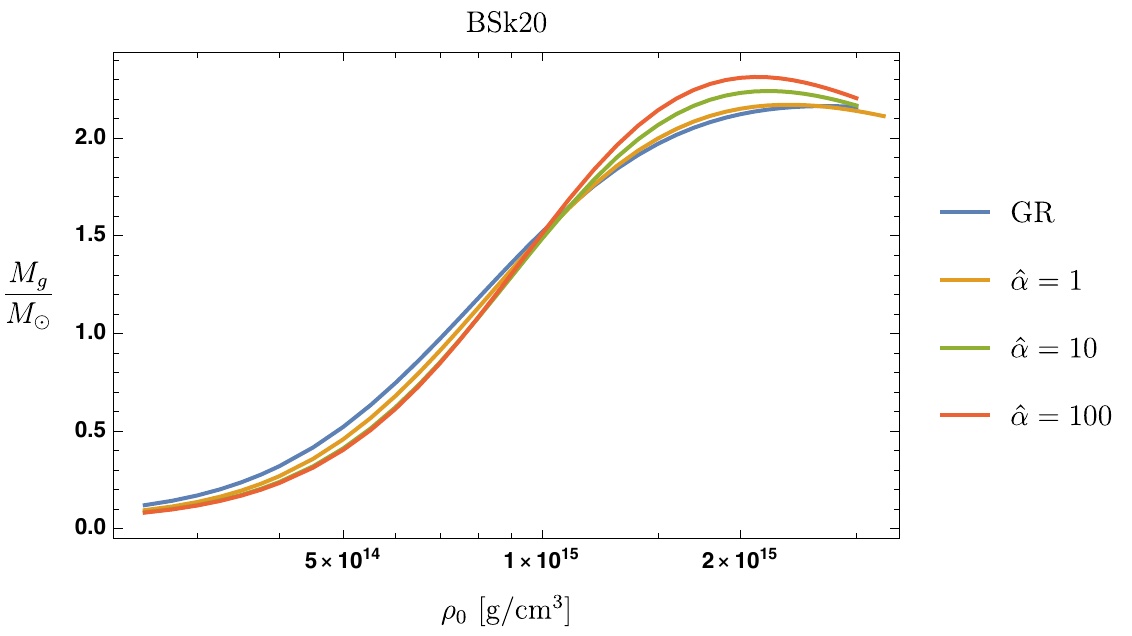}
	\caption{Same as in Figure \ref{Figure: BSk19 curves}, but for the BSk20 EoS.} \label{Figure: BSk20 curves}
\end{figure}

The f\mbox{}irst comment is that the M--R curves and the M--$\r_{_{0}}$ curves relative to the BSk equations of state share the same qualitative features with those relative to the Sly. In particular, for any def\mbox{}inition of gravitational mass, the mass increases monotonically with $\r_{_{0}}$ and there is an intermediate mass range where $dr_{\star}/dM > 0$ (noteworthy, for the BSk20 EoS this range exists already in GR). Moreover, there exists a critical value $\r_{_{0}}^{c}$ of the central density (which depends on the def\mbox{}inition of mass under consideration, and weakly on $\a$) such that for $\r_{_{0}} < \r_{_{0}}^{c}$ the mass in $R^{2}$-gravity is lower than the mass in GR with the same central density, while the opposite happens for $\r_{_{0}} > \r_{_{0}}^{c}$. This is apparent from the Figure \ref{Figure: BSk19 curves}, relative to the BSk19 EoS, and from the Figure \ref{Figure: BSk20 curves}, relative to the BSk20 EoS.

\subsubsection{Quantitative comparison}

A more ef\mbox{}fective comparison is provided by focusing on the features considered in Tables \ref{tab: M max Sly} and \ref{tab: delta M Sly}, that is the value of the maximum mass and the mass interval for which $dr_{\star}/dM > 0\,$. The analogous results for the BSk equations of state are given respectively in Tables \ref{tab: M max BSk19} and \ref{tab: delta M BSk19}, for the BSk19 EoS, and in Tables \ref{tab: M max BSk20} and \ref{tab: delta M BSk20}, for the BSk20 EoS.

\begin{table}[htb!]
\centering
\[
\begin{array}{cccccc}
\toprule
\hat{\a} \quad & \quad M^{\textup{max}}_{\textup{g}} \quad & \quad M^{\textup{max}}_{\r} \quad & \quad M^{\textup{max}}_{\textup{p}} \quad & \quad M^{\textup{max}}_{\textup{s}} \quad & \quad M^{\textup{max}}_{\textup{n}} \quad \\
\midrule
\textsc{gr} & 1.86 & 1.86 & 2.58 & 1.86 & 1.86 \\
1 & 1.88 & 1.89 & 2.59 & 1.80 & 1.70 \\
10 & 1.95 & 2.03 & 2.70 & 1.77 & 1.91 \\
100 & 2.01 & 2.14 & 2.81 & 1.78 & 2.01 \\
\bottomrule
\end{array}
\]
\caption{Maximum masses for GR and for $\hat{\a} = 1$, $10$, $100$, using the BSk19 EoS. Solar mass units are employed.}
\label{tab: M max BSk19}
\end{table}

\begin{table}[htb!]
\centering
\[
\begin{array}{cccccc}
\toprule
\hat{\a} \quad & \quad M^{\textup{max}}_{\textup{g}} \quad & \quad M^{\textup{max}}_{\r} \quad & \quad M^{\textup{max}}_{\textup{p}} \quad & \quad M^{\textup{max}}_{\textup{s}} \quad & \quad M^{\textup{max}}_{\textup{n}} \quad \\
\midrule
\textsc{gr} & 2.16 & 2.16 & 3.05 & 2.16 & 2.16 \\
1 & 2.17 & 2.18 & 3.01 & 2.09 & 1.95 \\
10 & 2.24 & 2.32 & 3.12 & 2.06 & 2.19 \\
100 & 2.31 & 2.45 & 3.25 & 2.07 & 2.30 \\
\bottomrule
\end{array}
\]
\caption{Same as in Table \ref{tab: M max BSk19}, but for the BSk20 EoS.}
\label{tab: M max BSk20}
\end{table}

Let us start commenting on the maximum gravitational mass. On a general ground, we can see that the maximum mass is higher with the BSk20 EoS than with the Sly EoS, and is lower with the BSk19 EoS. Apart from this, the behaviour of $M^{\textup{max}}$ displays evident similarities for the three equations of state. At f\mbox{}ixed $\hat{\a}$, the values of $M^{\textup{max}}$ are ordered as follows
\beq \label{hierarchy}
M^{\textup{max}}_{\textup{p}} > M^{\textup{max}}_{\r} > M^{\textup{max}}_{\textup{g}} > M^{\textup{max}}_{\textup{s}} \quad ,
\eeq
and for every value of $\hat{\a}$ we have $M^{\textup{max}}_{\textup{g}} > M^{\textup{max}}_{GR}$ and $M^{\textup{max}}_{\textup{s}} < M^{\textup{max}}_{GR}\,$. The analysis of section \ref{subsubsec: definitions of mass} is compatible with these results, and provides a theoretical explanation about why these properties are shared by all the equations of state considered here. It may indeed suggest these properties to be quite general. The value of $M^{\textup{max}}_{\textup{n}}$ on the other hand does not obey a clear ordering in relation to the other def\mbox{}initions of mass, although for all the equations of state considered it gets very close to $M^{\textup{max}}_{\textup{g}}$ when $\hat{\a} = 100\,$.

Regarding the dependence of $M^{\textup{max}}$ on $\hat{\a}$ when $1 \leq \hat{\a} \leq 100 \,$, note that for all the equations of state under consideration the values of $M^{\textup{max}}_{\textup{g}}$, $M^{\textup{max}}_{\r}$, $M^{\textup{max}}_{\textup{p}}$ and $M^{\textup{max}}_{\textup{n}}$ increase with increasing $\hat{\a}$, while the value of $M^{\textup{max}}_{\textup{s}}$ doesn't. Moreover, focusing on $M^{\textup{max}}_{\textup{g}}$, $M^{\textup{max}}_{\r}$ and $M^{\textup{max}}_{\textup{p}}$, the approximate logarithmic behaviour $M^{\textup{max}}(\hat{\a}) = M^{\textup{max}}(1) + k \, \log_{10} (\hat{\a})$ for $1 \leq \hat{\a} \leq 100$ (discussed in section \ref{subsubsec: Numerical values}) still holds with the BSk equations of state (although the agreement is not so good for $M^{\textup{max}}_{\r}$ with the BSk19). Intriguingly, as far as we can say from the study of the Sly, BSk19 and BSk20, the value of $k$ seems to depend very little (if at all) on the EoS.

\begin{table}[htb!]
\centering
\[
\begin{array}{cccccc}
\toprule
\hat{\a} & \D M_{\textup{g}} & \D M_{\r} & \D M_{\textup{p}} & \D M_{\textup{s}} & \D M_{\textup{n}} \\
\midrule
1 \quad & \quad 0.6 - 1.0 \quad & \quad 0.6 - 1.0 \quad & \quad 0.7 - 1.1 \quad & \quad 0.6 - 0.9 \quad & \quad 0.5 - 0.8 \\
10 \quad & \quad 0.4 - 1.2 \quad & \quad 0.4 - 1.3 \quad & \quad 0.5 - 1.5 \quad & \quad 0.3 - 1.0 \quad & \quad 0.4 - 1.3 \\
100 \quad & \quad 0.4 - 1.4 \quad & \quad 0.4 - 1.5 \quad & \quad 0.4 - 1.7 \quad & \quad 0.3 - 1.0 \quad & \quad 0.5 - 1.5 \\
\bottomrule
\end{array}
\]
\caption{Mass interval where $dr_{\!\star}/dM > 0\,$, using the BSk19 EoS. Solar mass units are employed, and the error is $\pm 0.1\,$.}
\label{tab: delta M BSk19}
\end{table}

\begin{table}[htb!]
\centering
\[
\begin{array}{cccccc}
\toprule
\hat{\a} & \D M_{\textup{g}} & \D M_{\r} & \D M_{\textup{p}} & \D M_{\textup{s}} & \D M_{\textup{n}} \\
\midrule
\textsc{gr} \quad & \quad 0.6 - 1.0 \quad & \quad 0.6 - 1.0 \quad & \quad 0.7 - 1.5 \quad & \quad 0.6 - 1.0 \quad & \quad 0.6 - 1.0 \\
1 \quad & \quad 0.5 - 1.5 \quad & \quad 0.5 - 1.5 \quad & \quad 0.5 - 1.8 \quad & \quad 0.4 - 1.4 \quad & \quad 0.4 - 1.2 \\
10 \quad & \quad 0.4 - 1.6 \quad & \quad 0.4 - 1.7 \quad & \quad 0.4 - 2.0 \quad & \quad 0.3 - 1.4 \quad & \quad 0.4 - 1.6 \\
100 \quad & \quad 0.4 - 1.8 \quad & \quad 0.4 - 2.0 \quad & \quad 0.4 - 2.3 \quad & \quad 0.3 - 1.4 \quad & \quad 0.5 - 1.9 \\
\bottomrule
\end{array}
\]
\caption{Same as in Table \ref{tab: delta M BSk19}, but for the BSk20 EoS.}
\label{tab: delta M BSk20}
\end{table}

Let us now pass to the intermediate mass range where $dr_{\!\star}/dM > 0\,$. Also in this case the behaviour of the interval $\D M$ displays evident similarities for the three equations of state. It is in fact apparent that $\D M$ becomes wider as $\hat{\a}$ increases, without shifting signif\mbox{}icantly its position. To be more specif\mbox{}ic, with the only exception of $\D M_{\textup{n}}\,$, the lower extreme of the interval decreases for $1 \leq \hat{\a} \leq 10$ and saturates for $10 \leq \hat{\a} \leq 100\,$, and its ``saturated'' value seems not to depend on the EoS. Furthermore, focusing on $\D M_{\textup{g}}$, $\D M_{\r}$ and $\D M_{\textup{p}}$, the lower extreme of the interval saturates to the \emph{same} value. The higher extreme of the interval, on the other hand, is monotonically increasing with increasing $\hat{\a}$, with the only exception of $\D M_{\textup{s}}\,$. Its value, dif\mbox{}ferently from the lower extreme, depends both on the def\mbox{}inition of mass and on the equation of state. 

\subsubsection{Comments on degeneracy}

Overall, the answer to the question whether there is degeneracy between the EoS and the def\mbox{}initions of gravitational mass seems to depend at least in part on what we exactly mean by saying that there is degeneracy. For example, it is well known (and our analysis conf\mbox{}irms it) that the maximum mass in general changes when the EoS changes. From our analysis it seems conceivable that the maximum mass correspondent to dif\mbox{}ferent def\mbox{}initions of mass can be made to have the same value by suitably changing the equation of state. Therefore, if our idea of degeneracy concerns only the maximum mass, and more specif\mbox{}ically only two def\mbox{}initions of gravitational mass, then we can say that there is degeneracy.

However, we may not be concerned only with the maximum value of two def\mbox{}initions of gravitational mass, but we may be interested also with the relation between the various def\mbox{}initions of mass, and with several other features of the M--R curves. From this latter point of view, our analysis gives indications that there are similarities/regularities in the features of the families of M--R curves which cannot be averted by changing the equation of state. An example is the hierarchy (\ref{hierarchy}) of the values of the maximum mass calculated with dif\mbox{}ferent def\mbox{}initions. This is especially true of the def\mbox{}initions $M_{\textup{g}}$, $M_{\r}$ and $M_{\textup{p}}$, while $M_{\textup{s}}$ and $M_{\textup{n}}$ have at times peculiar behaviours, depending on the feature under consideration.

Of course our analysis, having considered only three equations of state, cannot give def\mbox{}inite answers on this point but can merely give indications. To settle this point a thorough investigation, scanning a large number of equations of state, would be needed, but this may well be the subject of a separate publication.

\section{Discussion and conclusions}
\label{sec: Discussion and conclusions}

In the previous sections we discussed several possible def\mbox{}initions of gravitational mass for a static and spherically symmetric star in $R^{2}$ gravity, and unambiguously proved, by numerical means, that these def\mbox{}initions are indeed quantitatively dif\mbox{}ferent. This conf\mbox{}irms that, in modif\mbox{}ied theories of gravity, when speaking of the mass of a star it is not possible to avoid specifying which def\mbox{}inition is used, and caution has to be taken when estimating the properties of a star from observational data, especially when information about a star is obtained combining dif\mbox{}ferent observational techniques.

As we manifestly declared, our numerical results were obtained for values of $\a$ which are not compatible with observations. Our declared aim, however, was to use $R^{2}$-gravity as a proxy to study the issue related to the def\mbox{}inition of gravitational mass, due to its simplicity. It is fact known that modifying GR always introduces new degrees of freedom, and often introduces new characteristic scales. $R^{2}$-gravity is a very convenient proxy because it is simple yet nontrivial: it introduces only one extra DOF, whose potential is very simple (being just a mass term), and only one new scale (the mass $m$ or, in terms of length, the range $m^{-1}$). Choosing to work with values of $\a$ which span a wider domain than that allowed by observations has also advantages: it permits to appreciate more clearly the dif\mbox{}ference between the various def\mbox{}initions, and to perceive better how this dif\mbox{}ference depends on $\a\,$. For this reason, we believe our numerical results remain relevant, because document a phenomenon which is likely to be much more general than the particular model we considered.

\subsection{General considerations}

At f\mbox{}irst sight, the idea of having several inequivalent def\mbox{}initions of gravitational mass is surprising (disturbing, even). However, it is important to ref\mbox{}lect on the fact that in Newtonian gravity and in GR we use the concept of mass to characterise several a priori dif\mbox{}ferent things, such as for example the motion of test bodies in dif\mbox{}ferent locations in space (e.g.\ close to the star/far from the star), the gravitational redshift of radiation, the energy content of a star. We are so used to GR and Newton's theory that it is easy to take for granted that the ef\mbox{}fect of the gravitational f\mbox{}ield outside a static and spherically symmetric star can be described by a unique gravitational charge. The usefulness of using a specif\mbox{}ic model (here, $R^{2}$-gravity) to investigate a general problem shows up already here, since the study of static and spherically symmetric conf\mbox{}igurations in the weak f\mbox{}ield limit permits to show explicitly that the external gravitational f\mbox{}ield depends on \emph{two} gravitational charges.

Indeed we would like to reverse the perspective, and propose that, if we admit that a theory of gravity may contain others degrees of freedom, apart from those of the massless and spin-2 graviton of GR, and new characteristic scales, it is natural that a unique gravitational charge fails to describe the external gravitational f\mbox{}ield. It is just that the behaviour of the gravitational f\mbox{}ield is more complex. As the example of $R^{2}$-gravity shows, the crux of the problem is that the presence of matter excites the gravitational potentials and at the same time the extra DOF, which in turn couples to the gravitational potentials. In the process, the characteristic scale associated to the extra DOF remains imprinted in the behaviour of the gravitational potentials, along with the characteristic of the source. From this point of view, it is GR and Newton's theory that are special in their simplicity.

In our opinion, it is not that fruitful to try to establish which one is \emph{the} def\mbox{}inition of gravitational mass, and which are ancillary def\mbox{}initions. Choosing one def\mbox{}inition over the others would in some sense be like forcing the richer phenomenology of modif\mbox{}ied gravity into the conceptual structure of GR. We feel that it is better to live with the fact that gravity is more complex when you modify GR, and accept that dif\mbox{}ferent def\mbox{}initions of mass just describe dif\mbox{}ferent aspects of the gravitational f\mbox{}ield. On the other hand we deem important that every def\mbox{}inition of mass be tightly linked to specif\mbox{}ic observable phenomena, in continuity with the Newtonian's idea behind the introduction of gravitational mass.

\subsection{Comments on some specif\mbox{}ic def\mbox{}initions}

This of course does not mean that all def\mbox{}initions are equally useful, in practice. Regarding the motion of test bodies in the star's gravitational f\mbox{}ield, it is probably quite generic in MTG (at least in those with new characteristic scales) that the spacetime outside a star may be made up of dif\mbox{}ferent domains. In some of these domains Newtonian gravity is not reproduced, while in others it is, but the value of the ef\mbox{}fective Newtonian mass felt by test bodies (as well as the value of the local PPN parameters) vary from one domain to the other. In such a situation, to each domain where Newtonian gravity is reproduced we may assign an ef\mbox{}fective gravitational mass according to Kepler's third law
\beq \label{1-2-3 law}
M_{\textup{ef\mbox{}f}} = \o^{2} \, a^{3} \quad ,
\eeq
where $\o$ is the angular frequency ($\o = 2 \pi/T$, where $T$ is the period) and $a$ is the semi-major axis of the elliptical orbit.

On the other hand, we don't feel the def\mbox{}inition (\ref{stellar mass}) of mass $M_{\textup{s}}$ to be as useful as the others. First of all, its link to observable phenomena is weak. The idea of treating separately the extra DOF $\z$ inside and outside the star therefore seems dictated mainly by the desire to keep the concept of gravitational mass inside GR's conceptual framework. Secondly, it relies on the possibility of localising the energy of the extra DOF inside the star's surface, which in turn relies on the possibility of associating to $\z$ a (conserved) stress-energy tensor in a unique way. While a heuristic candidate for $\mcal{T}_{\m}{}^{\n}$ has been proposed, it is worthwhile to remind that, in the original fourth-order theory, the extra DOF is just a part of the gravitational f\mbox{}ield (it is essentially the curvature). To put the def\mbox{}inition of $M_{\textup{s}}$ on a f\mbox{}irm basis, it would be advisable to embed its def\mbox{}inition in a general analysis of the concept of energy of the gravitational f\mbox{}ield in the context of the fourth order theory.

\section*{Acknowledgments}
OFP thanks the Alexander von Humboldt foundation for funding and the Institute for Theoretical Physics of the Heidelberg University for kind hospitality. OFP wishes to thank Eduardo Grossi for his help in improving the Mathematica code employed for the research presented in this paper. OFP, POB and TM acknowledges partial f\mbox{}inancial support from CAPES (Brazil). This study was f\mbox{}inanced in part by the \emph{Coordena\c{c}\~ao de Aperfei\c{c}oamento de Pessoal de N\'ivel Superior} - Brazil (CAPES) - Finance Code 001. FS acknowledges partial f\mbox{}inancial support from CNPq and FAPES. SEJ acknowledges f\mbox{}inancial support from UFES in the occasion of a visit to the Astrophysics, Cosmology and Gravitation group.

\section*{Declarations of interest}
None


\begin{thebibliography}{99}

\bibitem{Demorest 2010}
P.B.~Demorest, T.~Pennucci, S.M.~Ransom, M.S.E.~Roberts and J.W.T.~Hessels,
\newblock {\em A two-solar-mass neutron star measured using Shapiro delay},
\newblock Nature \textbf{467} (2010) 1081,
\newblock \texttt{arXiv: 1010.5788 [astro-ph.HE]}.
\newblock {DOI: 10.1038/nature09466}.

\bibitem{Antoniadis 2013}
J.~Antoniadis et al.,
\newblock {\em A massive pulsar in a compact relativistic binary},
\newblock Science \textbf{340} (2013) 6131,
\newblock {\texttt{arXiv:1304.6875 [astro-ph.HE]}},
\newblock {DOI: 10.1126/science.1233232}.

\bibitem{TheLIGOScientific:2017qsa}
B.P.\ Abbott et al (LIGO Scientif\mbox{}ic, Virgo),
\newblock {\em GW170817: Observation of Gravitational Waves from a Binary Neutron Star Inspiral},
\newblock Phys.\ Rev.\ Lett.\ \textbf{119}, 16 (2017) 161101,
\newblock {\texttt{arXiv:1710.05832 [gr-qc]}},
\newblock {DOI: 10.1103/PhysRevLett.119.161101}.

\bibitem{Abbott:2017xzu}
B.P.\ Abbott et al,
\newblock {\em A gravitational-wave standard siren measurement of the Hubble constant} (collaborations: LIGO Scientif\mbox{}ic, Virgo, 1M2H, Dark Energy Camera GW-E, DES, DLT40, Las Cumbres Observatory, VINROUGE, MASTER)
\newblock Nature \textbf{551} (2017) 85,
\newblock {\texttt{arXiv:1710.05835 [astro-ph.CO]}},
\newblock {DOI: 10.1038/nature24471}.

\bibitem{LIGOScientific:2019fpa}
B.P.\ Abbott et al,
\newblock {\em Tests of General Relativity with the Binary Black Hole Signals from the LIGO-Virgo Catalog GWTC-1} (collaborations: LIGO Scientif\mbox{}ic, Virgo), (2019)
\newblock {\texttt{arXiv:1903.04467 [gr-qc]}}.

\bibitem{TheLIGOScientific:2016src}
B.P.\ Abbott et al,
\newblock {\em Tests of general relativity with GW150914} (collaborations: LIGO Scientif\mbox{}ic, Virgo)
\newblock Phys.\ Rev.\ Lett.\ \textbf{116} no.22 (2016) 221101,
\newblock {\texttt{arXiv:1602.03841 [gr-qc]}},
\newblock {DOI: 10.1103/PhysRevLett.116.221101}
\newblock [Erratum: {\em Phys.\ Rev.\ Lett.\ }\textbf{121} no.12 (2018) 129902,
\newblock {DOI: 10.1103/PhysRevLett.121.129902}].

\bibitem{Abbott:2018wiz}
B.P.\ Abbott et al,
\newblock {\em Properties of the binary neutron star merger GW170817} (collaborations: LIGO Scientif\mbox{}ic, Virgo)
\newblock Phys.\ Rev.\ \textbf{X 9} no.1 (2019) 011001,
\newblock {\texttt{arXiv:1805.11579 [gr-qc]}},
\newblock {DOI: 10.1103/PhysRevX.9.011001}.

\bibitem{Abbott:2018exr}
B.P.\ Abbott et al,
\newblock {\em GW170817: Measurements of neutron star radii and equation of state} (collaborations: LIGO Scientif\mbox{}ic, Virgo)
\newblock Phys.\ Rev.\ Lett.\ \textbf{121} no.16 (2018) 161101,
\newblock {\texttt{arXiv:1805.11581 [gr-qc]}},
\newblock {DOI: 10.1103/PhysRevLett.121.161101}.

\bibitem{Monitor:2017mdv}
B.P.\ Abbott et al,
\newblock {\em Gravitational Waves and Gamma-rays from a Binary Neutron Star Merger: GW170817 and GRB 170817A} (collaborations: LIGO Scientif\mbox{}ic, Virgo, Fermi-GBM, INTEGRAL)
\newblock Astrophys.\ J.\ \textbf{848} no.2 (2017) L13,
\newblock {\texttt{arXiv:1710.05834 [astro-ph.HE]}},
\newblock {DOI: 10.3847/2041-8213/aa920c}.

\bibitem{Yazadjiev:2014cza}
S.S.~Yazadjiev, D.D.~Doneva, K.D.~Kokkotas and K.V.~Staykov,
\newblock {\em Non-perturbative and self-consistent models of neutron stars in R-squared gravity},
\newblock JCAP \textbf{1406} (2014) 003,
\newblock {\texttt{arXiv:1402.4469 [gr-qc]}},
\newblock {DOI: 10.1088/1475-7516/2014/06/003}.

\bibitem{Yazadjiev:2015xsj}
S.S.~Yazadjiev and D.D.~Doneva,
\newblock {\em Comment on "The Mass-Radius relation for Neutron Stars in $f(R)$ gravity" by S. Capozziello, M. De Laurentis, R. Farinelli and S. Odintsov},
\newblock {\texttt{arXiv:1512.05711 [gr-qc]}}.

\bibitem{Capozziello:2015yza}
S.~Capozziello, M.~De Laurentis, R.~Farinelli and S.D.~Odintsov,
\newblock {\em Mass-radius relation for neutron stars in f(R) gravity},
\newblock Phys.\ Rev.\ \textbf{D 93} 2 (2016) 023501,
\newblock {\texttt{arXiv:1509.04163 [gr-qc]}},
\newblock {DOI: 10.1103/PhysRevD.93.023501}.

\bibitem{Starobinsky:1980te}
A.A.\ Starobinsky,
\newblock {\em A New Type of Isotropic Cosmological Models Without Singularity},
\newblock Phys.\ Lett.\ \textbf{B 91} (1980) 99--102,
\newblock {DOI: 10.1016/0370-2693(80)90670-X}.

\bibitem{starobinskynonsingular}
A.A.\ Starobinsky, 
\newblock {\em Nonsingular Model of the Universe with the Quantum-Gravitational de Sitter Stage and its Observational Consequences}
\newblock {in Quantum Gravitation. Quantum Theory of Gravitation} \textbf{1} (1982) 58--72.

\bibitem{Martin:2013nzq}
J.~Martin, C.~Ringeval, R.~Trotta and V.~Vennin,
\newblock {\em The Best Inf\mbox{}lationary Models After Planck},
\newblock JCAP \textbf{03} (2014) 039,
\newblock {\texttt{arXiv:1312.3529 [astro-ph.CO]}},
\newblock {DOI: 10.1088/1475-7516/2014/03/039}.

\bibitem{Amendola:2006we}
L.~Amendola, R.~Gannouji, D.~Polarski and S.~Tsujikawa,
\newblock {\em Conditions for the cosmological viability of f(R) dark energy models},
\newblock Phys.\ Rev.\ \textbf{D 75} (2007) 083504,
\newblock {\texttt{arXiv:[gr-qc/0612180]}},
\newblock {DOI: 10.1103/PhysRevD.75.083504}.

\bibitem{Akrami:2018odb}
Y.~Akrami et al (Planck collaboration),
\newblock {\em Planck 2018 results. X. Constraints on inf\mbox{}lation}, (2018)
\newblock {\texttt{arXiv:1807.06211 [astro-ph.CO]}}.

\bibitem{Bertotti:2003rm}
B.\ Bertotti, L.\ Iess and P.\ Tortora,
\newblock {\em A test of general relativity using radio links with the Cassini spacecraft},
\newblock Nature \textbf{425} (2003) 374,
\newblock {DOI: 10.1038/nature01997}.

\bibitem{Will:2014kxa}
W.M.\ Clif\mbox{}ford,
\newblock {\em The Confrontation between General Relativity and Experiment},
\newblock Living Rev.\ Rel. \textbf{17} no.1 (2014) 4,
\newblock {\texttt{arXiv:1403.7377 [gr-qc]}},
\newblock {DOI: 10.12942/lrr-2014-4}.

\bibitem{Sotiriou:2008rp}
T.P.\ Sotiriou and V.\ Faraoni,
\newblock {\em f(R) Theories Of Gravity},
\newblock Rev.\ Mod.\ Phys.\ \textbf{82} (2010) 451--497,
\newblock {\texttt{arXiv:0805.1726 [gr-qc]}},
\newblock {DOI: 10.1103/RevModPhys.82.451}.

\bibitem{DeFelice:2010aj}
A.\ De Felice and S.\ Tsujikawa,
\newblock {\em f(R) theories},
\newblock Living Rev.\ Rel.\ \textbf{13} (2010) 3,
\newblock {\texttt{arXiv:1002.4928 [gr-qc]}},
\newblock {DOI: 10.12942/lrr-2010-3}.

\bibitem{Nojiri:2010wj}
S.~Nojiri and S.D.~Odintsov,
\newblock {\em Unif\mbox{}ied cosmic history in modif\mbox{}ied gravity: from F(R) theory to Lorentz non-invariant models},
\newblock Phys.\ Rept.\ \textbf{505} (2011) 59--144,
\newblock {\texttt{arXiv:1011.0544 [gr-qc]}},
\newblock {DOI: 10.1016/j.physrep.2011.04.001}.

\bibitem{Nojiri:2017ncd}
S.~Nojiri, S.D.~Odintsov and V.K.~Oikonomou,
\newblock {\em Modif\mbox{}ied Gravity Theories on a Nutshell: Inf\mbox{}lation, Bounce and Late-time Evolution},
\newblock Phys.\ Rept.\ \textbf{692} (2017) 1--104,
\newblock {\texttt{arXiv:1705.11098 [gr-qc]}},
\newblock {DOI: 10.1016/j.physrep.2017.06.001}.

\bibitem{Teyssandier:1983zz}
P.~Teyssandier and P.~Tourrenc,
\newblock {\em The Cauchy problem for the R+R**2 theories of gravity without torsion},
\newblock J.\ Math.\ Phys.\ \textbf{24} (1983) 2793,
\newblock {DOI: 10.1063/1.525659}.

\bibitem{Kase:2019dqc}
R.~Kase and S.~Tsujikawa,
\newblock {\em Neutron stars in $f(R)$ gravity and scalar-tensor theories}, (2019)
\newblock {\texttt{arXiv:1906.08954 [gr-qc]}}.

\bibitem{Sbisa:2018rem}
F.~Sbis\`a, O.F.~Piattella and S.E.~Jor\'as,
\newblock {\em Pressure ef\mbox{}fects in the weak-f\mbox{}ield limit of f(R) = R + alpha $R^2$ gravity},
\newblock Phys.\ Rev.\ \textbf{D 99} 10 (2019) 104046,
\newblock {\texttt{arXiv:1811.01322 [gr-qc]}},
\newblock {DOI: 10.1103/PhysRevD.99.104046}.

\bibitem{Resco:2016upv}
M.~Aparicio Resco, \'{A}.~de la Cruz-Dombriz, F.J.~Llanes-Estrada and V.~Zapatero Castrillo,
\newblock {\em On neutron stars in $f(R)$ theories: small radii, large masses and large energy emitted in a merger},
\newblock Phys.\ Dark\ Univ.\ \textbf{13} (2016) 147--161,
\newblock {\texttt{arXiv:1602.03880 [gr-qc]}},
\newblock {DOI: 10.1016/j.dark.2016.07.001}.

\bibitem{Astashenok:2017dpo}
A.V.~Astashenok, S.D.~Odintsov and \'{A}.~de la Cruz-Dombriz,
\newblock {\em The realistic models of relativistic stars in $f(R) = R + \a R^{2}$ gravity},
\newblock Class.\ Quant.\ Grav.\ \textbf{34} 20 (2017) 205008,
\newblock {\texttt{arXiv:1704.08311 [gr-qc]}},
\newblock {DOI: 10.1088/1361-6382/aa8971}.

\bibitem{Chiba:2003ir}
T.~Chiba,
\newblock {\em 1/R gravity and scalar-tensor gravity},
\newblock Phys.\ Lett.\ \textbf{B 575} (2003) 1--3,
\newblock {\texttt{arXiv:[astro-ph/0307338]}},
\newblock {DOI: 10.1016/j.physletb.2003.09.033}.

\bibitem{Wald 1984}
R.M.~Wald,
\newblock \emph{{General Relativity}},
\newblock  {The University of Chicago Press}, Chicago and London (1984).

\bibitem{Misner 1973}
C.W.~Misner, K.S.~Thorne and J.A.~Wheeler,
\newblock {\em {Gravitation}},
\newblock W. H. Freeman, San Francisco (1973).

\bibitem{Capozziello:2002rd}
S.~Capozziello,
\newblock {\em Curvature quintessence},
\newblock Int.\ J.\ Mod.\ Phys.\ \textbf{D 11} (2002) 483--492,
\newblock {\texttt{arXiv:[gr-qc/0201033]}},
\newblock {DOI: 10.1142/S0218271802002025}.

\bibitem{Capozziello:2006uv}
S.~Capozziello, V.F.~Cardone and A.~Troisi,
\newblock {\em Dark energy and dark matter as curvature ef\mbox{}fects},
\newblock {JCAP} \textbf{0608} (2006) 001,
\newblock {\texttt{arXiv:[astro-ph/0602349]}},
\newblock {DOI: 10.1088/1475-7516/2006/08/001}.

\bibitem{Van Paradijs 1979}
J.\ Van Paradijs,
\newblock {\em Possible observational constraints on the mass-radius relation of neutron stars},
\newblock Astrophys.\ J.\ \textbf{234} (1979) 609--611,
\newblock {DOI: 10.1086/157535}.

\bibitem{Astashenok:2013vza}
A.V.\ Astashenok, S.\ Capozziello and S.D.\ Odintsov,
\newblock {\em Further stable neutron star models from f(R) gravity},
\newblock JCAP {\bf 1312} (2013) 040,
\newblock {\texttt{arXiv:1309.1978 [gr-qc]}},
\newblock {DOI: 10.1088/1475-7516/2013/12/040}.

\bibitem{Astashenok:2014pua}
A.V.\ Astashenok, S.\ Capozziello and S.D.\ Odintsov,
\newblock {\em Maximal neutron star mass and the resolution of the hyperon puzzle in modif\mbox{}ied gravity},
\newblock Phys.\ Rev.\ {\bf D 89} no.10 (2014) 103509,
\newblock {\texttt{arXiv:1401.4546 [gr-qc]}},
\newblock {DOI: 10.1103/PhysRevD.89.103509}.

\bibitem{Astashenok:2014dja}
A.V.\ Astashenok, S.\ Capozziello and S.D.\ Odintsov,
\newblock {\em Nonperturbative models of quark stars in $f$(R) gravity},
\newblock Phys.\ Lett.\ {\bf B 742} (2015) 160--166,
\newblock {\texttt{arXiv:1412.5453 [gr-qc]}},
\newblock {DOI: 10.1016/j.physletb.2015.01.030}.

\bibitem{Nari:2018aqs}
N.\ Nari and M.\ Roshan,
\newblock {\em Compact stars in Energy-Momentum Squared Gravity},
\newblock Phys.\ Rev.\ {\bf D 98} no.2 (2018) 024031,
\newblock {\texttt{arXiv:1802.02399 [gr-qc]}},
\newblock {DOI: 10.1103/PhysRevD.98.024031}.

\bibitem{Chabanat:1997un}
E.~Chabanat, P.~Bonche, P.~Haensel, J.~Meyer and R.~Schaef\mbox{}fer,
\newblock {\em A Skyrme parametrization from subnuclear to neutron star densities. 2. Nuclei far from stablities},
\newblock Nucl.\ Phys.\ \textbf{A 635} (1998) 231--256,
\newblock {DOI: 10.1016/S0375-9474(98)00570-3},
\newblock [Erratum: Nucl.\ Phys.\ \textbf{A 643} (1998) 441,
\newblock {DOI: 10.1016/S0375-9474(98)00180-8}].

\bibitem{Douchin:2001sv}
F.~Douchin and P.~Haensel,
\newblock {\em A unif\mbox{}ied equation of state of dense matter and neutron star structure},
\newblock Astron.\ Astrophys.\ \textbf{380} (2001) 151,
\newblock {\texttt{arXiv:[astro-ph/0111092]}},
\newblock {DOI: 10.1051/0004-6361:20011402}.

\bibitem{Haensel:2004nu}
P.~Haensel and A.Y.~Potekhin,
\newblock {\em Analytical representations of unif\mbox{}ied equations of state of neutron-star matter},
\newblock Astron.\ Astrophys.\ \textbf{428} (2004) 191--197,
\newblock {\texttt{arXiv:[astro-ph/0408324]}},
\newblock {DOI: 10.1051/0004-6361:20041722}.

\bibitem{Staykov:2014mwa}
K.V.~Staykov, D.D.~Doneva, S.S.~Yazadjiev and K.D.~Kokkotas, 
\newblock {\em Slowly rotating neutron and strange stars in $R^2$ gravity},
\newblock {JCAP} \textbf{1410} (2014) 006,
\newblock {\texttt{arXiv:1407.2180 [gr-qc]}},
\newblock {DOI: 10.1088/1475-7516/2014/10/006}.

\bibitem{Yazadjiev:2015zia}
S.S.~Yazadjiev, D.D.~Doneva and K.D.~Kokkotas,
\newblock {\em Rapidly rotating neutron stars in R-squared gravity},
\newblock Phys.~Rev.~\textbf{D 91}, 8 (2015) 084018,
\newblock {\texttt{arXiv:1501.04591 [gr-qc]}},
\newblock {DOI: 10.1103/PhysRevD.91.084018}.

\bibitem{Sorkin:1981jc}
R.\ Sorkin,
\newblock {\em A Criterion for the onset of instability at a turning point},
\newblock Astrophys.\ J.\ \textbf{249} (1981) 254--257,
\newblock {DOI: 10.1086/159282}.

\bibitem{Green:2013ica}
S.R.\ Green, J.S.\ Schif\mbox{}frin and R.M.\ Wald,
\newblock {\em Dynamic and Thermodynamic Stability of Relativistic, Perfect Fluid Stars},
\newblock Class.\ Quant.\ Grav.\ \textbf{31} (2014) 035023,
\newblock {\texttt{arXiv:1309.0177 [gr-qc]}},
\newblock {DOI: 10.1088/0264-9381/31/3/035023}.

\bibitem{Ozel:2016oaf}
F.\ \"Ozel and P.\ Freire,
\newblock {\em Masses, Radii, and the Equation of State of Neutron Stars},
\newblock Ann.\ Rev.\ Astron.\ Astrophys.\ {\bf 54} (2016) 401--440,
\newblock {\texttt{arXiv:1603.02698 [astro-ph.HE]}},
\newblock {DOI: 10.1146/annurev-astro-081915-023322}.

\bibitem{Potekhin:2013qqa}
A.Y.\ Potekhin, A.F.\ Fantina, N.\ Chamel, J.M.\ Pearson and S.\ Goriely,
\newblock {\em Analytical representations of unif\mbox{}ied equations of state for neutron-star matter},
\newblock Astron.\ Astrophys.\ {\bf 560} (2013) A48,
\newblock {\texttt{arXiv:1310.0049 [astro-ph.SR]}},
\newblock {DOI: 10.1051/0004-6361/201321697}.





\end{thebibliography}
\end{document}